\documentclass[12pt]{iopart}

\usepackage[english]{babel}
\usepackage{graphicx}
\usepackage{cite}
\usepackage{subfigure}

\newcommand{\I}{\mathcal{I}}
\newcommand{\0}{\underline{\;\,}\,}
\newcommand{\1}{\underline{\bullet}\,}

\begin{document}

\title[Dynamical Transition in the TASEP]{Dynamical Transition in the 
Open-boundary Totally Asymmetric Exclusion Process} 
\author{A. Proeme, R. A. Blythe and M. R. Evans}
\address{SUPA, School of Physics and Astronomy, University of
Edinburgh, Mayfield Road, Edinburgh EH9 3JZ, United Kingdom}
\eads{\mailto{arno.proeme@ed.ac.uk},
\mailto{r.a.blythe@ed.ac.uk},
\mailto{m.evans@ed.ac.uk}}

\begin{abstract}
We revisit the totally asymmetric simple exclusion process with open boundaries (TASEP),
focussing on the recent discovery by de~Gier and Essler that the model has a dynamical transition along a nontrivial line in the phase diagram. This line coincides neither with any change in the steady-state properties of the TASEP, nor the corresponding line predicted by domain wall theory.  We provide numerical evidence that the TASEP indeed has a dynamical transition along the de~Gier--Essler line, finding that the most convincing evidence was obtained from Density Matrix Renormalisation Group (DMRG) calculations.  By contrast, we find that the dynamical transition is rather hard to see in direct Monte Carlo simulations of the TASEP.  We furthermore discuss in general terms scenarios that admit a distinction between static and dynamic phase behaviour.

\medskip
\noindent Date: October 26, 2010
\end{abstract}

\maketitle

\section{Introduction}
\label{sec:intro}
As a rare example of a class of exactly-solvable, nonequilibrium interacting
particle systems, asymmetric simple exclusion processes of various types have found
favour with researchers in statistical mechanics, mathematical physics
and probability theory \cite{Liggett99,Derrida98,GM06,BE07}. These systems comprise
hard-core particles hopping in a preferred direction on a one-dimensional lattice and have been
used to describe systems as diverse as traffic flow \cite{Popkov01},
the dynamics of ribosomes \cite{Macdonald68} and molecular motors
\cite{AGP99,KL03},
the flow of hydrocarbons through a zeolite pore \cite{Chatterjee09},
the growth of a fungal filament \cite{Sugden07}, and in queueing
theory \cite{Martin07,EFM09}. The physical interest lies mainly in the rich
phase behaviour that arises as a direct consequence of being driven
away from equilibrium.

Of particular interest are the open boundary cases. Here the
system can be thought of as being placed between two boundary
reservoirs, generally of different densities.  The two reservoirs
enforce a steady current across the lattice, and therewith a
nonequilibrium steady state.  It was first argued by Krug
\cite{Krug91} that as the boundary densities are varied nonequilibrium
phase transitions may occur in which steady state bulk
quantities---such as the mean current or density---exhibit
nonanalyticities.  Such phase transitions were seen explicitly in
first a mean-field approximation \cite{Macdonald68,DDM92} and then in
the exact solution \cite{DEHP93,Schutz93} of the totally asymmetric exclusion process with open boundaries,
hereafter abbreviated as TASEP, and of related processes \cite{BE07}.  Since then arguments have been developed
to predict the phase diagram of more general one-dimensional driven
diffusive systems \cite{Popkov99,HKPS01,MB05}.

\begin{figure}
  \centering
  \subfigure[]{\label{fig:spd}%
    \includegraphics[width=0.45\linewidth]{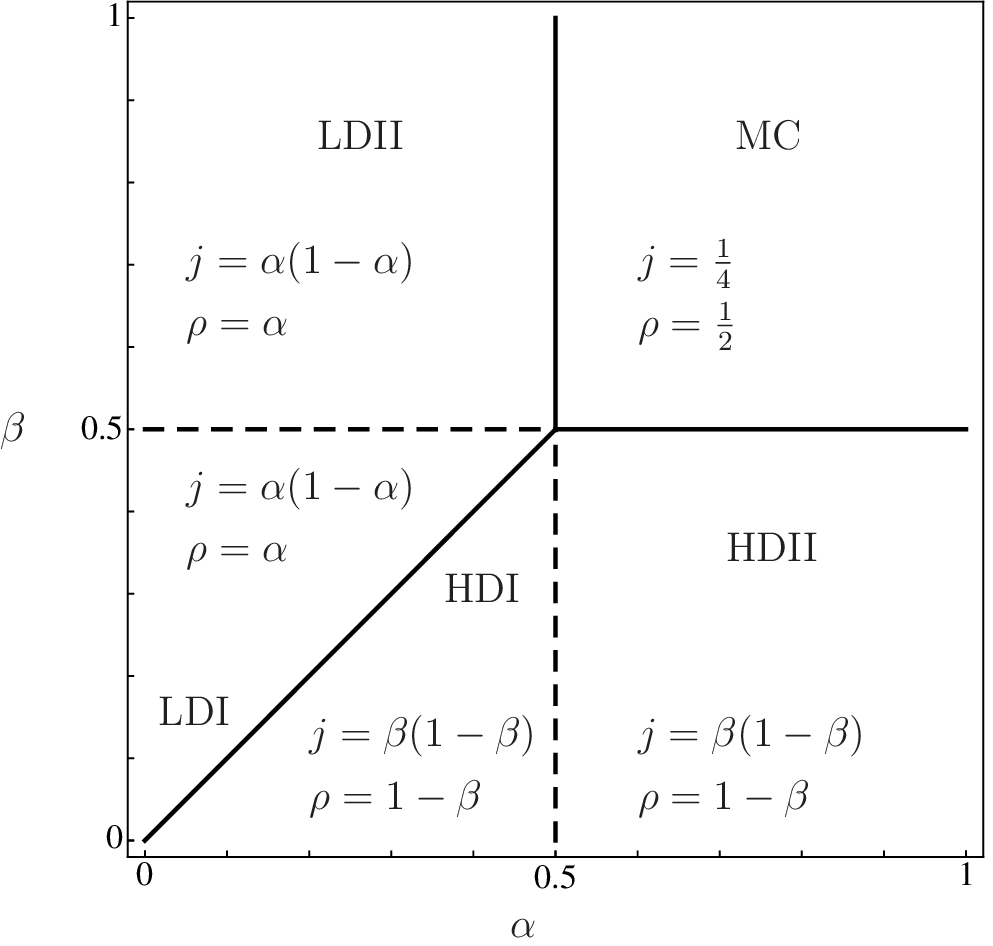}}
  \subfigure[]{\label{fig:dynpd}%
    \includegraphics[width=0.45\linewidth]{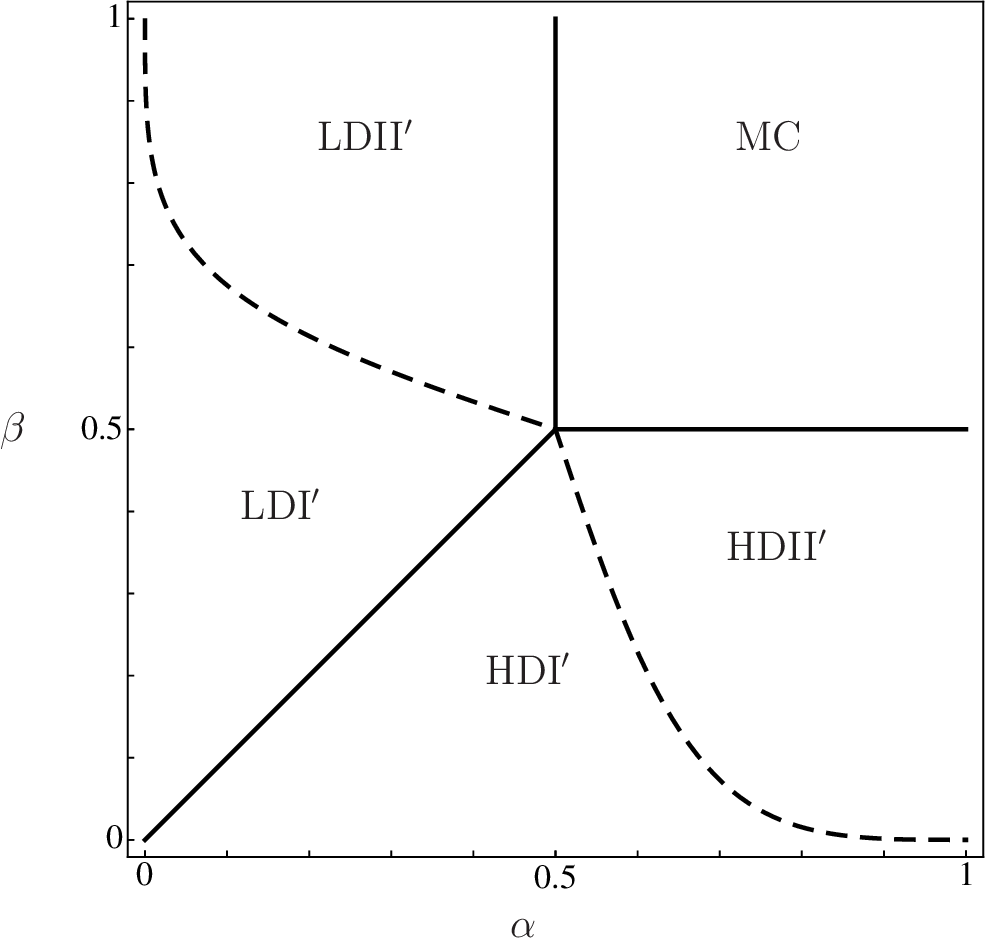}} 
  \caption{ \subref{fig:spd} Static and \subref{fig:dynpd} dynamic phase diagrams for the TASEP. Solid lines indicate thermodynamic phase transitions at which the current and bulk density are nonanalytic. Dotted lines indicate subphase boundaries. In the static case, the density profile a finite distance from the boundary is nonanalytic. In the dynamic case, the longest relaxation time exhibits a nonanalyticity.}
\end{figure}

Our interest in this work is in the distinction between two phase diagrams for the TASEP:
one of which characterises the steady-state behaviour, and the other the dynamics.  The static phase diagram \cite{DEHP93,Schutz93} is shown in
Fig.~\ref{fig:spd}  where the left boundary density is $\alpha$ and the right is $1-\beta$.
There are three possible phases in the steady state: a low density (LD) phase where the bulk density is
controlled by the left boundary and is equal to $\alpha$; a high
density (HD) phase where the bulk density is controlled by the right
boundary and is equal to $1-\beta$; a maximal current (MC) phase where the
bulk density is $\frac{1}{2}$.  The high and low density phases are
further divided into subphases (e.g. LD1, LDII)  according to the functional form of
the spatial decay of the density profile to the bulk value near the
non-controlling  boundary.  In these subphases the lengthscale over which the decay is
observable remains finite as the system size is increased. This change
at the subphase boundaries in the form of the decay
is thus not a true phase transition in the thermodynamic sense.

More recently, de~Gier and Essler have performed an exact
analysis of the ASEP's dynamics \cite{deGier05, deGier06, deGier08}
and derived the the longest relaxation times of the system by
calculating the gap in the spectrum using the Bethe ansatz. 
The analysis builds on directly related work on the XXZ spin chain with nondiagonal
boundary fields \cite{Neopmechiel03a, Neopmechiel03b}.
The dynamical phase diagram they obtain, in which phases are associated
with different functional forms of the longest relaxation time, is illustrated in figure
\ref{fig:dynpd}.  The same phases (high density, low density and
maximal current) are found as in the static phase diagram, however, a
hitherto unexpected \emph{dynamical transition} line which subdivides the
low density and high density phases is now apparent. This line which
we shall refer to as the de Gier--Essler (dGE) line replaces the subphase
boundaries at $\alpha=1/2, \beta <1/2$ $\beta=1/2, \alpha <1/2$ in the
static phase diagram and, for example, subdivides the low density region into LDI$'$, LDII$'$.

Although there is nothing to rule out the prediction of dGE of a
dynamical transition at a distinct location to any static transition,
the result came as something of a surprise. This is because the static
phase diagram had been successfully interpreted in terms of an
effective, dynamical theory thought to be relevant for late-time
dynamics, referred to as domain wall theory (DWT).  We shall review
DWT more fully in Section~\ref{subsec:DWT}.  For the moment we note
that DWT correctly predicts the static subphase boundary and the associated
change in the decay of the density profile, but does not predict the
dGE line.  Therefore, it was previously thought that a dynamical transition also
occurred at the subphase boundary and initial calculations of
relaxation times appeared to be consistent with this
\cite{DS00,Nagy02}.

Our primary goal in this work is to establish numerically that a dynamical
transition does indeed occur along the dGE line rather than at the
static subphase boundary.  We used two distinct approaches.  First, we carried
out direct Monte Carlo simulations of the TASEP dynamics and identified
the dominant transient in three time-dependent observables.  We find that these
simulations are consistent with a dynamical transition at the dGE line
but are not sufficiently accurate to rule out the scenario of the
dynamical transition occurring at the subphase boundary.

We instead turn to a Density Matrix Renormalisation Group (DMRG) approach
\cite{DMRGBook} to acquire convincing evidence that a dynamical transition
occurs at the dGE line. The DMRG is an approximate means to obtain the
lowest-lying energies and eigenstates of model Hamiltonians with short-range
interactions. Although originally developed in the context of the Hubbard
and Heisenberg models \cite{White92} (see also \cite{Schollwock05} for a
comprehensive review), it has also been applied to nonequilibrium processes
such as reaction-diffusion models \cite{Carlon99}.  In particular, it has
also been used to investigate the spectrum of the TASEP \cite{Nagy02}.
However, the dynamical transition was not evident from the data presented in
that work---perhaps because it had not been predicted at that time.  By
revisiting this approach, we obtain estimates of the longest relaxation time
that allow us to rule out the domain wall theory prediction for the
dynamical transition.

The paper is organised as follows. We begin in Section~\ref{sec:moddef}
by recalling the definition of the TASEP, and by reviewing in more detail the
static and dynamic phase diagrams discussed above. Then, in Section~\ref{sec:sims},
we present our Monte Carlo simulation data, followed by the DMRG results in Section~\ref{sec:dmrg}.  In Section~\ref{sec:discuss} we return to more general questions about the distinction between the static and dynamic phase diagram with particular reference to different theoretical approaches to the TASEP.  We conclude in Section~\ref{sec:conclude} with some open questions.

\section{Model definition and phase diagrams}
\label{sec:moddef}

Although we have alluded to the basic properties of the totally asymmetric simple exclusion
process (TASEP) in the introduction, in
the interest of a self-contained presentation we provide in this section a precise model
definition and full details of the static and dynamic phase diagrams.

\subsection{Model definition}

\begin{figure}
\begin{center}
\includegraphics[width=0.55\linewidth]{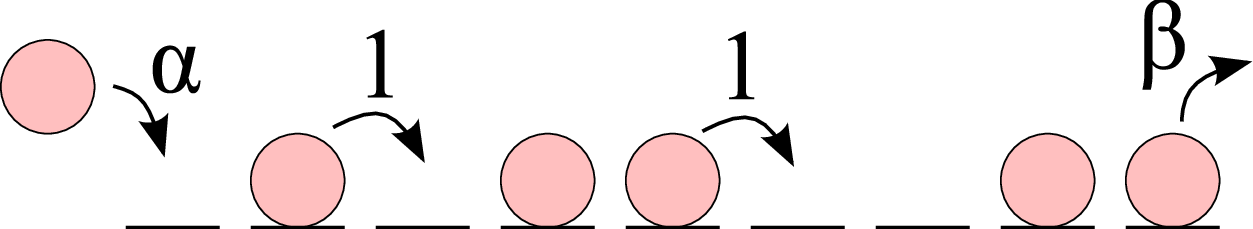}
\end{center}
\caption{\label{fig:asep} The dynamics of the totally asymmetric simple exclusion process (TASEP) with open boundaries.  Arrows show moves that may take place, and the labels indicate the corresponding stochastic rates.}
\end{figure}

The TASEP describes the biased diffusion of particles on a
one-dimensional lattice with $L$ sites. No more than one particle can
occupy a given site, and overtaking is prohibited. The stochastic
dynamics are expressed in terms of transition rates, for example a
particle on a site $i$ in the bulk hops to the right as a Poisson
process with rate $1$, but only if site $i+1$ is unoccupied:
At the left boundary only particle influx takes
place, with rate $\alpha$, and at the right boundary only particle
outflux takes place, with rate $\beta$, as shown in Fig.~\ref{fig:asep}. The corresponding
reservoir densities are $\alpha$ at the left and $1-\beta$ at the right.

\subsection{Static phase diagram}

As noted in the introduction, the TASEP static phase diagram can be
determined from, for example, exact expressions for the current and density
profile in the steady state \cite{DEHP93,Schutz93}.  The current takes
three functional forms according to whether it is limited by a slow insertion
rate (LD: $\alpha < \beta, \alpha < \frac{1}{2}$), by a slow exit rate
(HD: $\beta < \alpha, \beta < \frac{1}{2}$) or by the exclusion interaction
in the bulk (MC: $\alpha>\frac{1}{2}, \beta>\frac{1}{2}$).  This behaviour of the
current leads to the identification of the thermodynamic phases separated by
solid lines in Fig.~\ref{fig:spd}. Along these lines there are nonanalyticities
in both the current and the density far from either boundary.  The relevant expressions for these quantities are shown on Fig.~\ref{fig:spd}.

The density \emph{near} one of the boundaries is nonanalytic along the lines
$\alpha=\frac{1}{2}$ and $\beta=\frac{1}{2}$ in the HD and LD phases respectively. These
subphases are shown dotted in Fig.~\ref{fig:spd}.  Inspection of the functional form
of the density profile reveals that these are not true thermodynamic phase transitions,
in the sense that the deviation from the bulk value extends only a finite distance into the bulk,
and thus contributes only subextensively to the nonequilibrium analogue of a free energy (see e.g., \cite{BE02} for the definition of such a quantity).
To be more explicit, consider for example the behaviour
near the right boundary in the low-density phase. Here the bulk density is $\rho=\alpha$. The
mean occupancy of the lattice site positioned a distance $j$ from the
right boundary approaches in the thermodynamic limit $L\to\infty$ the
form \cite{Schutz93, DEHP93, BE07}
\begin{equation}
\label{eq:rhoLD}
\rho_{L-j} \sim \left\{ \begin{array}{l@{\qquad}l@{\qquad}l}
\alpha + c_{\rm I}(\beta) \left( \frac{\alpha(1-\alpha)}{\beta(1-\beta)} \right)^{j} & \alpha < \beta < \frac{1}{2} & \mbox{(LDI)}\\
\alpha + c_{\rm II}(\alpha, \beta) \frac{ [ 4 \alpha(1-\alpha) ]^{j} }{j^{3/2}} & \frac{1}{2} < \beta & \mbox{(LDII)}
\end{array}\right. \;.
\end{equation}
In these expressions, $c_{\rm I}$ and $c_{\rm II}$ are functions of the boundary
parameters that we leave unspecified here so as to focus on the lengthscale of the
exponential decay from the right boundary. As $\beta$ is increased from zero, the decay length increases until
 $\beta=\frac{1}{2}$. Then the decay length is constant, and the exponential is modulated by a power law.  The
density profile at the left boundary exhibits the same kind of nonanalyticity
in the high-density phase as $\alpha$ is increased through
$\frac{1}{2}$ as a consequence of the particle-hole symmetry,
$\rho_{i-1}(\alpha,\beta) = 1-\rho_{L-i}(\beta,\alpha)$, exhibited by
the model.

\subsection{Dynamical phase diagram}
\label{sec:dynpd}

The dynamic phase diagram is obtained by examining a quantity referred to as the \emph{gap} by de~Gier and Essler \cite{deGier05,deGier06,deGier08}. This is simply the largest nonzero eigenvalue of the transition matrix governing the continuous-time stochastic dynamics of the TASEP.  More formally, one starts with the master equation
\begin{equation} 
  \label{eq:me}
  \frac{{\rm d}}{{\rm d}t} |P(t) \rangle = M |P(t) \rangle \;,
\end{equation}
where the matrix $M$ encodes the transition rates and $|P(t)\rangle$ is the vector of probabilities for each configuration. Because $M$ is a stochastic matrix it is guaranteed \cite{Altenberg02} by the Perron-Frobenius theorem to have eigenvalues satisfying 
\begin{equation}
  \lambda_0 = 0 \, > \, {\rm Re}(\lambda_1) \,\geq\, {\rm Re}(\lambda_2) \geq \ldots \;.
\end{equation}
The spectrum corresponds to a set of relaxation times $\tau_i = -({\rm Re}\, \lambda_i)^{-1}$. The longest (non-infinite) relaxation time is $\tau_1$, and the associated eigenvalue $\lambda_1$ is the gap which we will henceforth denote by the symbol $\varepsilon$.  At long times the relaxation of any observable is expected to decay exponentially with a characteristic timescale $\tau_1$, a fact we will later use to estimate the gap from Monte Carlo simulations. 

The thermodynamic phase boundaries are found by identifying lines along which the gap vanishes, indicating the coexistence of two stationary eigenstates (phases).  The exact calculations of de~Gier and Essler \cite{deGier05,deGier06} show that, in the thermodynamic limit $L\to\infty$, the gap vanishes along the
line $\alpha=\beta<\frac{1}{2}$ that separates the HD and LD phases. The gap also vanishes in the entirety of the MC phase, reflecting the generic long-range (power-law) correlations seen in this phase \cite{DEHP93}. Thus at this level, the static and dynamic phase diagrams coincide.

Where they differ is in the subdivision of the high- and low-density phases\footnote{de~Gier and Essler \cite{deGier05,deGier06,deGier08} refer to these as ``massive'' phases by analogy with the quantum spin chains; we shall only use the terminology associated with the static phase diagram to avoid confusion.}, in which the gap remains finite in the limit $L\to\infty$.   There is a region, marked LDI$'$ and HDI$'$ on Fig.~\ref{fig:dynpd}, within which the gap assumes the asymptotic expression
 \begin{equation} 
    \label{eq:gap_MI}
    \varepsilon(L) = -\alpha - \beta + \frac{2}{(ab)^{\frac{1}{2}} + 1} - \frac{\pi^2}{ (ab)^{\frac{1}{2}} - (ab)^{-\frac{1}{2}}} L^{-2} + \mathcal{O}(L^{-3})
  \end{equation}
in which 
\begin{equation}
a = \frac{1-\alpha}{\alpha} \quad \mbox{and}\quad  b =  \frac{1-\beta}{\beta} \;.
\end{equation}
Within the low-density phase ($\alpha<\beta, \alpha<\frac{1}{2}$), this form of the gap applies for values of $\beta < \beta_c$ where
\begin{equation}  
  \label{eq:beta_c}
  \beta_c(\alpha) = \left[1 +\left(\frac{\alpha}{1-\alpha}\right)^{\frac{1}{3}}\right]^{-1} \;.
\end{equation}
Likewise, in the high-density phase ($\beta<\alpha, \beta<\frac{1}{2}$), the region within which the gap is given by (\ref{eq:gap_MI}) is bounded by $\alpha<\alpha_c$ where
\begin{equation}
  \label{eq:alpha_c}
  \alpha_c(\beta) = \left[1 +\left(\frac{\beta}{1-\beta}\right)^{\frac{1}{3}}\right]^{-1} \;.
\end{equation}
In the remainder of the low-density phase, $\alpha<\frac{1}{2}, \beta>\beta_c$ the gap takes the form
\begin{equation}
\label{eq:gap_MIIa}
    \varepsilon(L) = -\alpha - \beta_c + \frac{2}{(ab_c)^{\frac{1}{2}} + 1} - \frac{4\pi^2}{ (ab_c)^{\frac{1}{2}} - (ab_c)^{-\frac{1}{2}}} L^{-2} + \mathcal{O}(L^{-3})\;.
\end{equation}
Finally, we have by symmetry that when $\beta<\frac{1}{2}, \alpha>\alpha_c$,
\begin{equation} 
  \label{eq:gap_MIIb}
  \varepsilon(L) = -\alpha_c - \beta + \frac{2}{(a_cb)^{\frac{1}{2}} + 1} - \frac{4\pi^2}{ (a_cb)^{\frac{1}{2}} - (a_cb)^{-\frac{1}{2}}} L^{-2} + \mathcal{O}(L^{-3}) \;.
\end{equation}
In these expressions,
\begin{equation}
a_c = \frac{1-\alpha_c}{\alpha_c} \quad\mbox{and}\quad b_c = \frac{1-\beta_c}{\beta_c} \;.
\end{equation}
The boundaries between the dynamic subphases---the de~Gier--Essler (dGE) lines---are shown dotted in Fig.~\ref{fig:dynpd}. Note that the coefficient of the $L^{-2}$ term in this asymptotic expansion is discontinuous across the dynamical transition line.

\begin{figure}
  \centering 
  \includegraphics[width=0.7\linewidth]{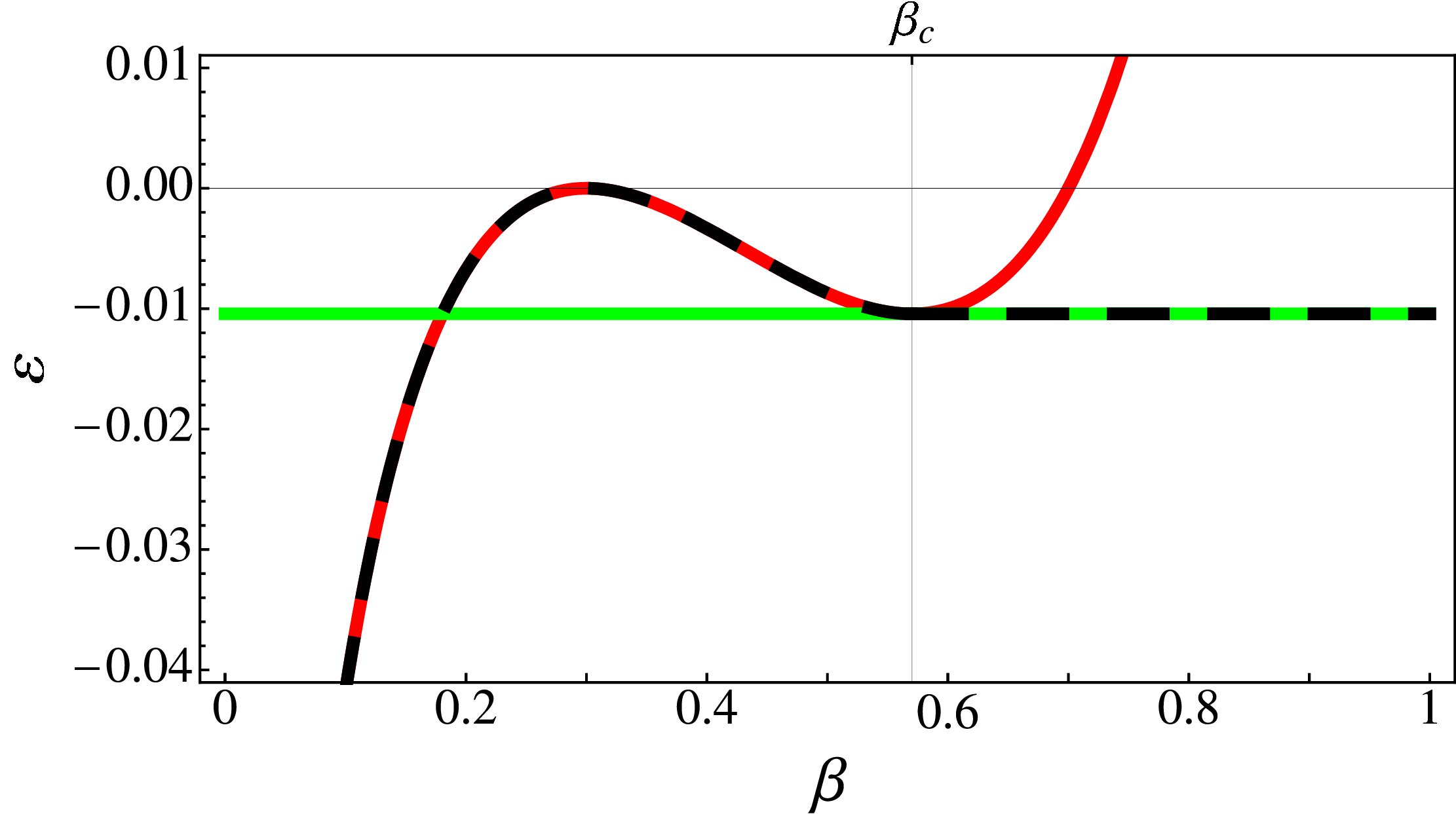}
  \caption{  \label{fig:deGierGapInf} Exact $L\to\infty$ gap (black, dashed) as a function of $\beta$ for $\alpha = 0.3$, described by the HDI$'$ equation~(\ref{eq:gap_MI}) (red) for $\beta < \beta_c\,$ ($\beta_c \approx 0.57$, indicated), and by equation~(\ref{eq:gap_MIIa}) (green) for $\beta > \beta_c$.}
\end{figure}

In order to illustrate the behaviour of the gap along the static and dynamic transition lines, we plot it as a
function of $\beta$ at some $\alpha < \frac{1}{2}$ in Fig.~\ref{fig:deGierGapInf}.  The most striking feature is
the constancy of the gap above the nontrivial critical point $\beta_c$.   Later in this work, we will use the constancy of the gap above a critical threshold as an empirical means to identify the dynamical transition point.

\subsection{Domain wall theory}
\label{subsec:DWT}

An alternative approach to study the TASEP dynamics is domain wall theory (DWT)~\cite{Kolomeisky98}. Although much simpler than the Bethe ansatz approach of de~Gier and Essler, it correctly predicts the static phase boundaries and the analytical form of the gap in a region of the phase diagram.  However, it predicts dynamic subphases that are distinct from those found by de~Gier and Essler, and that correspond to the static subphases.  In the numerical study that follows, it will be important to be able to distinguish between these two sets of predictions, and so we summarise DWT here.

\begin{figure}
\centering
\includegraphics[width=0.5\linewidth]{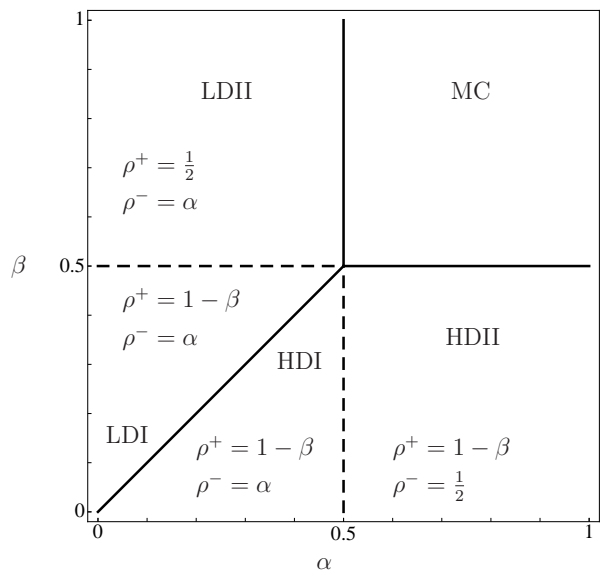}
\caption{\label{fig:DWTPhaseDiagram} The phase diagram obtained from domain wall theory.  $\rho^{+}$ and $\rho^{-}$ indicate the densities of the left and right domains.  In the domain wall theory, static and dynamic transitions coincide along the dashed lines.}
\end{figure}

The basis of this approach is to assume that collective relaxational dynamics
are effectively reducible to a single coordinate describing the position of an interface, or domain 
wall. The wall separates a domain of density $\rho^-$ and current $j^-$ to
the left from a domain of density $\rho^+$ and current $j^+ =
\rho^+(1-\rho^+)$ to the right. Each domain is taken to possess the
steady state characteristics imposed by the boundary reservoir on
that particular side of the wall, with $\rho^+$ and $\rho^-$
therefore as in Fig.~\ref{fig:DWTPhaseDiagram}. 
The effective theory is expected to be exact along the line $\alpha= \beta <1/2$ where
the exact  properties of the system  such as 
density profile \cite{DEHP93,Schutz93} and current fluctuations \cite{DEM95} are recovered.

Let us consider the case $\alpha <1/2$ and $\beta <1/2$. The motion of the wall is then described microscopically as a random walker  with left and right hopping rates $D^-$ and $D^+$  given respectively  by imposing mass conservation on the fluxes into and out of the wall \cite{DS00}:
\begin{equation}
\label{D+D-}
  D^- = \frac{j^-}{\rho^+ - \rho^-}= \frac{\alpha(1-\alpha)}{1-\alpha-\beta} \qquad  D^+ = \frac{j^+}{\rho^+ - \rho^-}
=\frac{\beta(1-\beta)}{1-\alpha-\beta},
\end{equation}
For  $\alpha > \beta$ the random walk is biased to the left
and in the stationary state the domain wall is localised at the left boundary
and the bulk density is given by $\rho^+=1-\beta$.
For  $\alpha < \beta$ the random walk is biased to the right
and in the stationary state the domain wall is localised at the right boundary
and the bulk density is given by $\rho^-=\alpha$. Thus the first order transition at $\alpha= \beta$ is correctly predicted.

Moreover, the gap in the resulting spectrum for large $L$ given \cite{DS00} is given by
\begin{equation}
  \label{eq:DWTGap}
  \varepsilon_{DWT}= -D^+ - D^- + 2\sqrt{D^+D^-}\left(1-\frac{\pi^2}{2L^2} + O(L^{-3}) \right).
\end{equation} 
Remarkably this expression is identical to (\ref{eq:gap_MI})
to order $1/L^2$. Thus  in the region  $\alpha<1/2, \beta <1/2$ DWT correctly predicts the gap.

When $\beta >1/2$ so that $\rho^+ = 1/2$, one can take $j^+$ in (\ref{D+D-}) to depend on the size $\ell$ of the right-hand domain. That is, one puts $j^+ \to j^+(\ell)$ equal to the stationary current in a TASEP of size $\ell$ in the maximal current phase. This implies  the large $\ell$ behaviour
\begin{equation}
j^+ \simeq \frac{1}{4}(1 + \frac{3}{2\ell})
\label{lpmc}
\end{equation}
This results in a modification of the density profile to an exponential spatial decay  modulated by a power law with power 3/2, similar to Eq. \ref{eq:rhoLD} (LDII).

\begin{figure}
\centering
\includegraphics[width=0.66\linewidth]{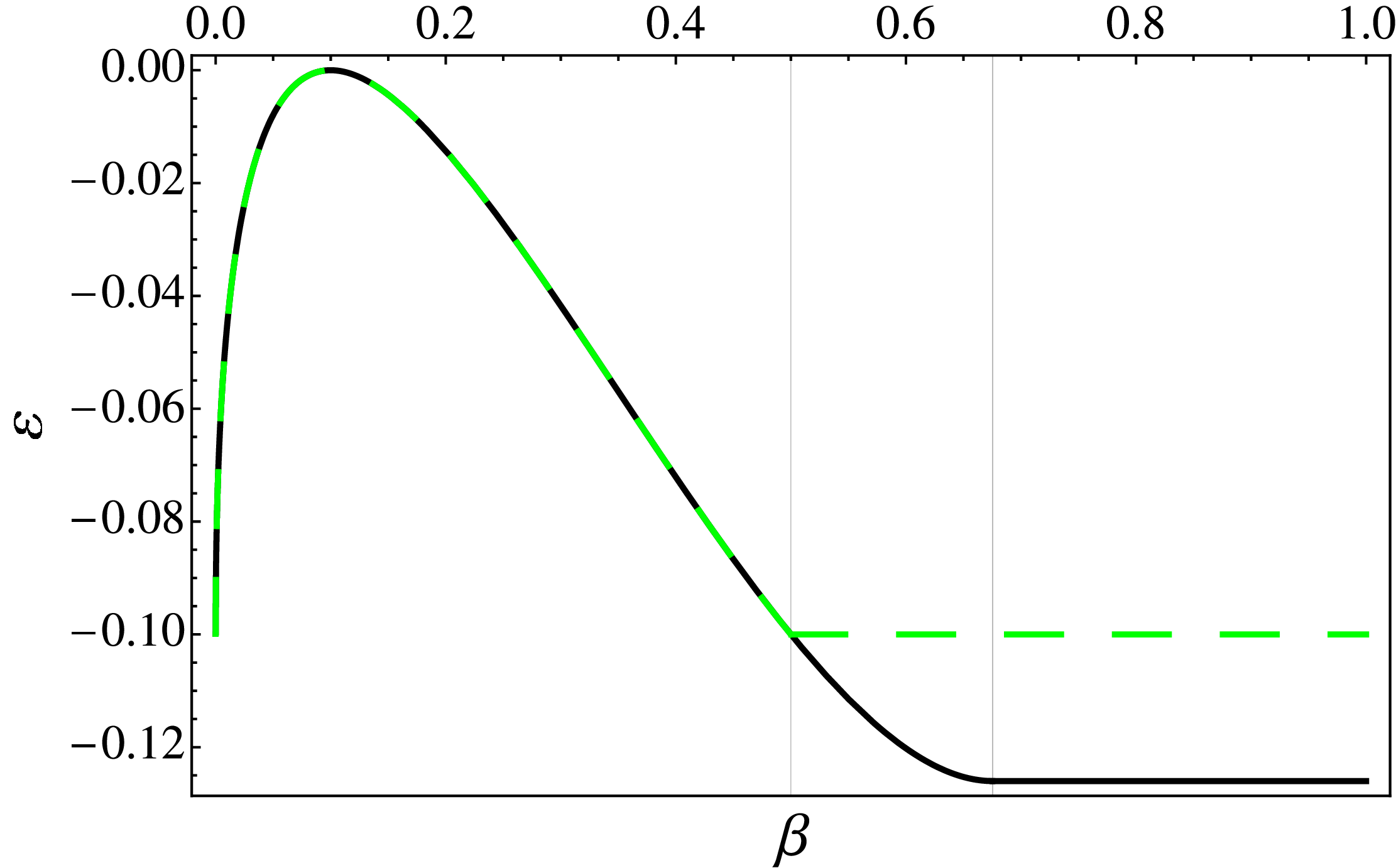}
\caption{\label{fig:GapComparisonDWTExact} Comparison of the gap for $L \to \infty$ from domain wall theory (dashed line) with the exact gap (solid line) at $\alpha = 0.1$ (for which $\beta_c \approx 0.675$).}
\end{figure}

In brief, the DWT is remarkably successful, correctly predicting the static phase diagram (including subphases), and the
exact thermodynamic gap function found by de~Gier and Essler in the region $\alpha<1/2$ and $\beta <1/2$. It differs in that the dynamic subphases are not given by the dGE lines, but the static subphase transition lines.  Thus, the region within which the gap is constant is different in the two theories, as indicated by Fig.~\ref{fig:GapComparisonDWTExact}.

\section{Evidence for the dynamical transition from Monte Carlo simulations}
\label{sec:sims}

We now describe attempts to measure the gap in Monte Carlo simulations of the TASEP, with a particular interest in distinguishing between the dGE and DWT predictions.  As noted previously, the ensemble average of any time-dependent observable ${\cal O}(t)$ should be dominated at late times by an exponential decay ${\rm e}^{-|\lambda_1| t}$ to its stationary value $\langle {\cal O} \rangle$. Recall that $\lambda_1$ is the largest nonzero eigenvalue of the matrix $M$ appearing in the master equation (\ref{eq:me}), and that all nonstationary eigenvalues have negative real part.  In principle, therefore, all one needs to do is pick an observable, and measure its late-time decay rate.  In practice, this is made difficult by the fact that \emph{all} decay modes may contribute to $\langle {\cal O}(t)\rangle$, and that by the time the higher modes have decayed away, the residual signal $\langle{\cal O}(t)\rangle - \langle {\cal O} \rangle$ may be extremely small and swamped by the noise.

Since we are interested in the time-dependence of an observable, we employ a \emph{continuous-time} (Gillespie) algorithm \cite{Gillespie77} to simulate the TASEP dynamics.  More precisely, we maintain a list of events (i.e., a particle hopping to the next site, or entering or leaving the system) that can take place given the current lattice configuration.  A particular event $i$ is then chosen with a probability proportional to its rate $\omega_i$ as specified in Section~\ref{sec:moddef}.  A time variable is then incremented by an amount chosen from an exponential distribution with mean $( \sum_i \omega_i )^{-1}$.  In this way, a member of the ensemble of all continuous-time trajectories of the TASEP dynamics from some prescribed initial condition is generated with the appropriate probability.

\subsection{Decay of total occupancy to stationarity}
\label{sec:decay}

We consider first the decay of the total number of particles on the lattice, $N(t)$, to its stationary value as measured by the function
\begin{equation}
\label{Rt}
R(t) = \frac{\langle N(t) \rangle - \langle N \rangle}{\langle N(0) \rangle - \langle N \rangle} \;.
\end{equation}
In this expression, angle brackets denote an average over an ensemble of initial conditions and over stochastic trajectories.  In each simulation run, an initial condition was constructed in which each site was independently occupied with a probability $p=1-\beta$ (in the LD phase) or $p=\alpha$ (in the HD phase). In the bulk, these densities then relax to the steady-state values displayed on Fig.~\ref{fig:spd}.

Once $R(t)$ has been sampled over multiple trajectories, the task is to identify a time window over which one can fit an exponential decay and therewith estimate a gap.  The start and end points of this windows are both crucial. If it starts too early, then one may expect contributions from subdominant transients (i.e., the decays at rate $\lambda_2, \lambda_3, \ldots$) to systematically skew the estimate of the gap.  If it ends too late, noise may instead dominate the estimate.

\begin{figure}
  \centering
  \subfigure[]{\label{fig:mu}%
    \includegraphics[height=4.2cm]{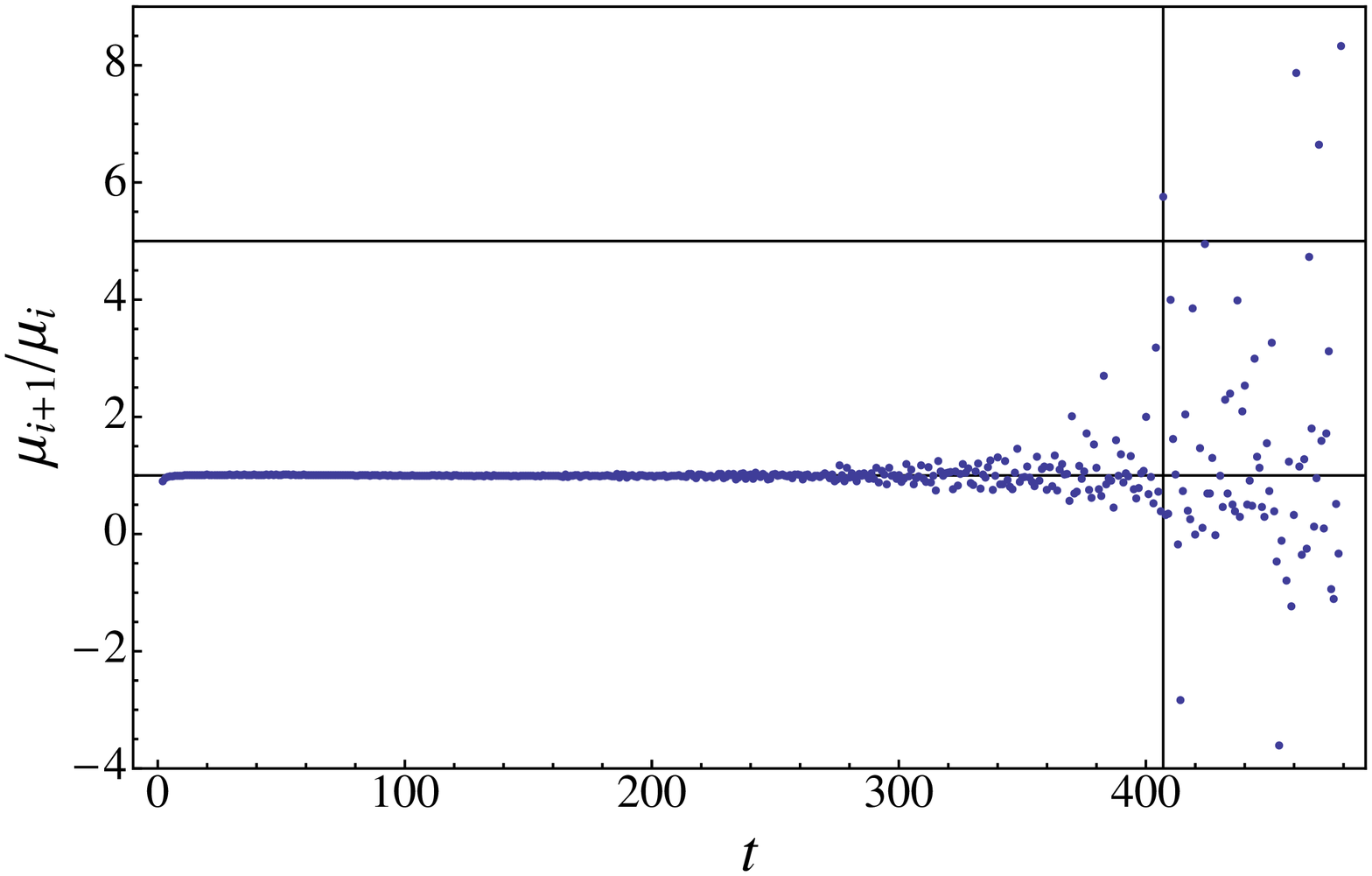}}
  \subfigure[]{\label{fig:R2mc}%
    \includegraphics[height=4.2cm]{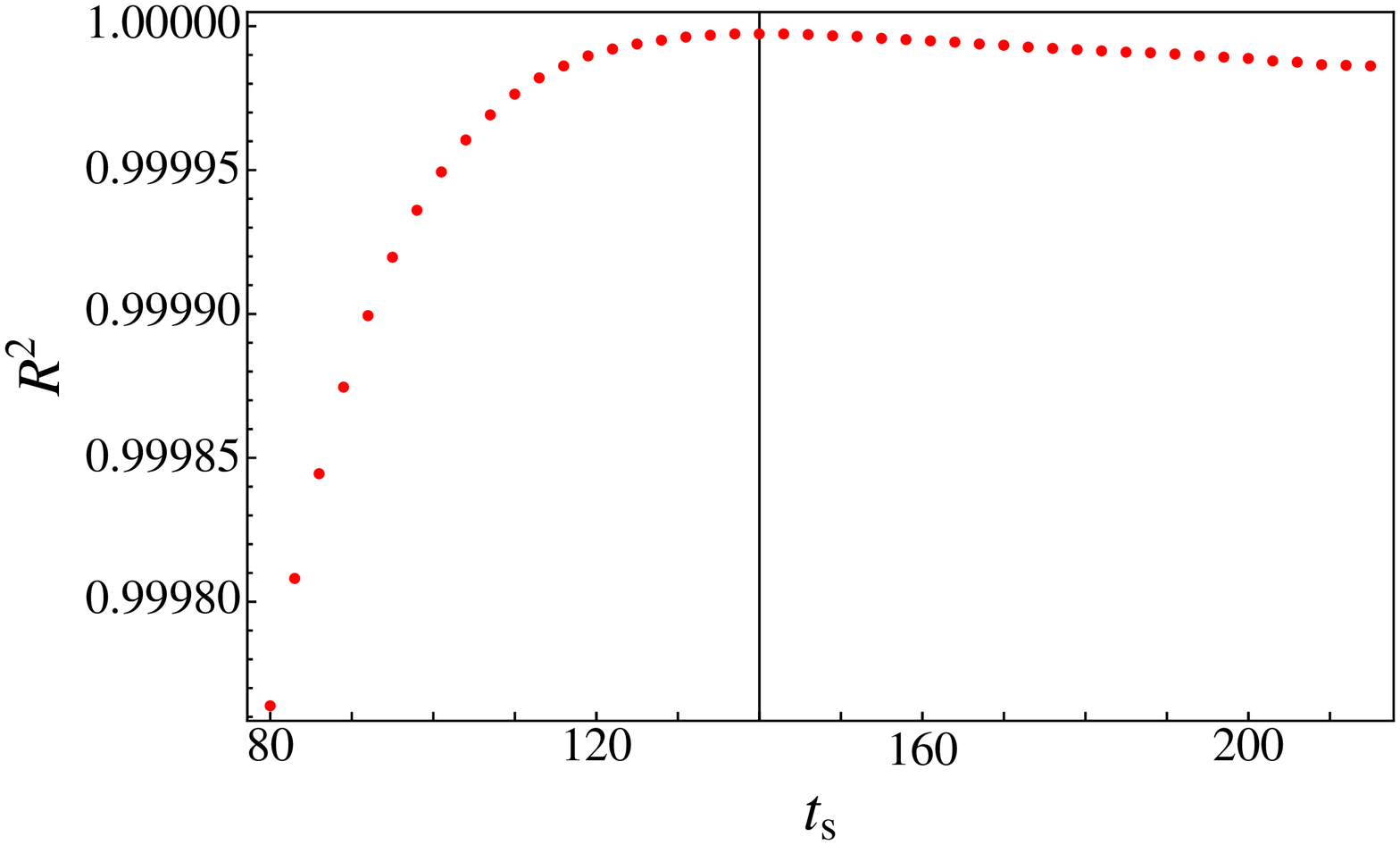}} 
  \caption{Criteria for choosing the range of data over which to fit a single exponential to the decay of the density to stationarity. \subref{fig:mu} The ratio of gradients $\mu_{i+1}/\mu_i$ which should be close to unity where a single exponential is a good fit. At large times, this ratio becomes noisy; the presence of extreme values (here, a magnitude larger than 5) is used to identify the onset of noise. \subref{fig:R2mc} The behaviour of ${\cal R}^2$ as a function of the start of the window. We choose this such that ${\cal R}^2$ is maximised.  In both cases, $L=50$, $\alpha=0.3$ and $\beta=0.64$.}
\end{figure}

The noise at the top end we handle by examining the behaviour of local decay rates
\begin{equation}
\mu_i = \frac{\ln R(t_{i+i}) - \ln R(t_{i})}{t_{i+1} - t_i}
\end{equation}
where $t_i$ and $t_{i+1}$ are successive time points at which $R(t)$ was sampled.  At late times, one should find $\mu_{i+1}/\mu_{i} \to 1$. Strong deviations from unity indicate the dominance of noise, and we rejected points after which the magnitude of this ratio exceeded a critical value. For our data sets, we found that $5$ was a suitable choice for this value: see Fig.~\ref{fig:mu} for an example.

The bottom end of the window was chosen by maximising a goodness-of-fit measure to a fit of the exponential $f(t) = a{\rm e}^{-\lambda t}$ to data points within the window.  We adopted the \emph{adjusted coefficient of determination} \cite{Ryan1997}
\begin{equation}
\label{R2}
{\cal R}^2 = 1 - \frac{\frac{1}{n-k}\sum_{i} (R(t_i) - f(t_i))^2}{\frac{1}{n-1}\sum_{i} (R(t_i) - \bar{R})^2} \;.
\end{equation}
as our goodness-of-fit measure, varying the start of the window until this quantity was maximised.  In this expression $\bar{R}$ is the arithmetic mean of $R(t)$ over the set of $n$ times $t_i$ falling within the window, and $k$ is the number of free parameters in the fit function $f(t)$, i.e., $k=2$.  This goodness-of-fit measure trades off the increased quality of the fit obtained by discarding data points against the increasing uncertainty in the fitting parameters that comes with the noisier data at the top end of the window.  We show in Fig.~\ref{fig:R2mc} an example of how the goodness-of-fit varies with the size of the window. We remark that the relative flatness of ${\cal R}^2$ above the optimal starting time can be ascribed to the fact that only a single decay mode remains in this region.  Although this procedure is a little cumbersome to apply to each data set, we found it preferable to fitting multiple exponentials to the data for $R(t)$, which leads to large uncertainties in the measured gap.

\begin{figure}
\begin{center}
\includegraphics[width=0.66\linewidth]{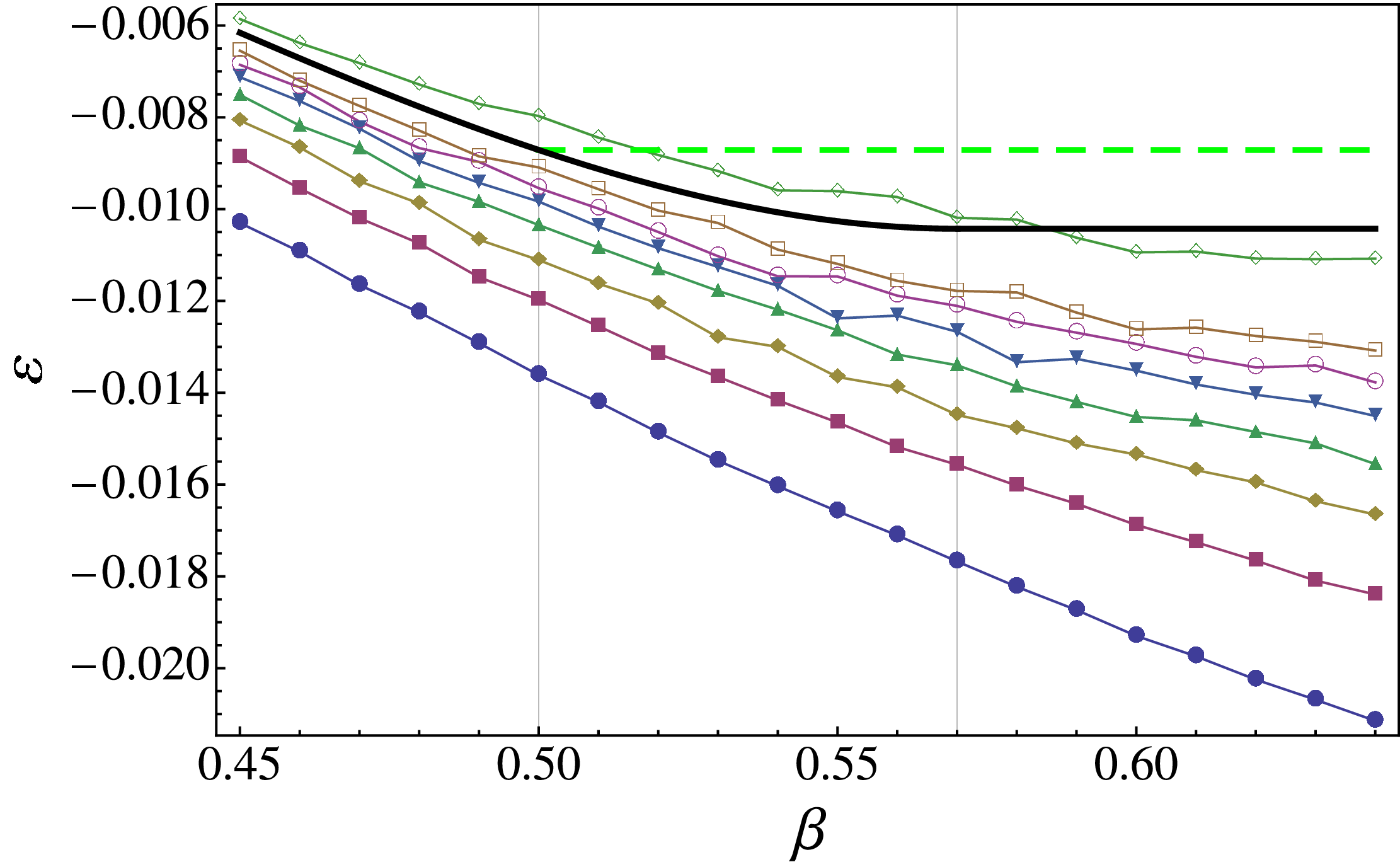}
\end{center}
\caption{\label{fig:mcgapfinite} The gap as a function of $\beta$ at $\alpha=0.3$ and system sizes $L$ ranging from $L=40$ to $L=150$ (from bottom to top) obtained from Monte Carlo simulations.  The $L\to\infty$ dGE (solid line) and DWT (dashed line) predictions are shown for comparison: even with the largest system $L=150$ simulated, one cannot distinguish between them numerically.}
\end{figure}

We show results in Fig.~\ref{fig:mcgapfinite}, in which the gap is plotted as a function of $\beta$ for fixed $\alpha=0.3$ for a range of system sizes.  At the smallest system sizes ($L=10$) data from Monte Carlo simulation are in agreement with results obtained through exact diagonalisation of the matrix $M$.  As the system size is increased beyond the sizes for which exact techniques remain tractable, the curves approach those predicted by dGE and DWT.  However, it is not possible to discern from the largest system simulated, $L=150$, which of these theories is more appropriate.

\begin{figure}
\begin{center}
\includegraphics[width=0.66\linewidth]{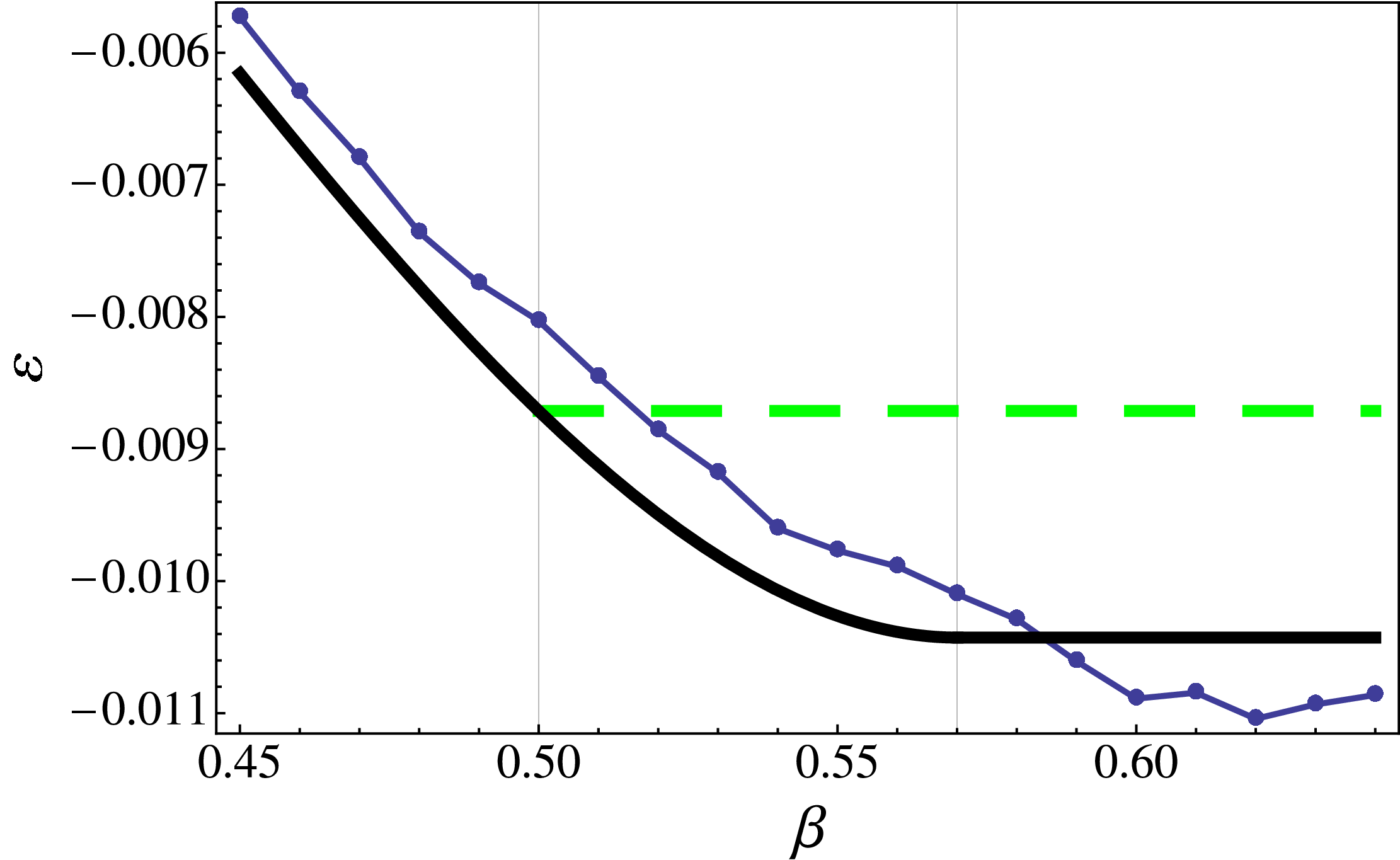}
\end{center}
\caption{\label{fig:mcgap} The gap extrapolated to $L=\infty$ from the data shown in Fig.~\ref{fig:mcgapfinite}.  Again the dGE (solid line) and DWT (dashed line) predictions are shown for comparison: here the data are potentially more compatible with the former than the latter.}
\end{figure}

We therefore attempt to extrapolate to the limit $L\to\infty$ by fitting a function of the form $\varepsilon + a_2 L^{-2} + a_3 L^{-3} + a_4 L^{-4}$ to estimates of the gap at finite system sizes $L$ but fixed $\alpha$ and $\beta$.  This particular fitting function was found to have the best goodness-of-fit (taking into account the varying number of parameters) out of those that were tried, and furthermore had the most uniform goodness-of-fit as a function of $\beta$.  The results of this extrapolation procedure are shown in Fig.~\ref{fig:mcgap}.  These data appear to exclude the functional form for the domain wall theory, although the uncertainties in the estimated gaps are such that further evidence is needed in order to confidently rule it out.

\subsection{Decay of other observables}

One possible means to reduce the uncertainty in the measured gap is to try different observables ${\cal O}(t)$ in the hope that contributions from subdominant decay modes---and in particular the second longest-lived mode that decays at rate $\lambda_2$---are reduced and thereby allow a more accurate estimate of $\lambda_1$.  We considered two candidates.  

The first was the autocorrelation function of the total occupancy, ${\cal O}(t) = N(t_0) N(t_0+t)$, where $t_0$ is some fixed time point in the steady state.  The analogue of the function $R(t)$ is
\begin{equation}
\chi(t) = \frac{\langle N(t_0+t)N(t_0) \rangle - \langle N \rangle^2}{\langle N^2 \rangle - \langle N \rangle^2}
\label{chi}
\end{equation}
where here all averages are taken in the steady state. Numerically, this quantity is slightly more convenient than $N(t)$, since one can obtain $\langle N(t_0+t)N(t_0) \rangle$ by sampling at different $t_0$ and $t$ from a single simulation run that has reached the steady state.

\begin{figure}
\begin{center}
\includegraphics[width=0.55\linewidth]{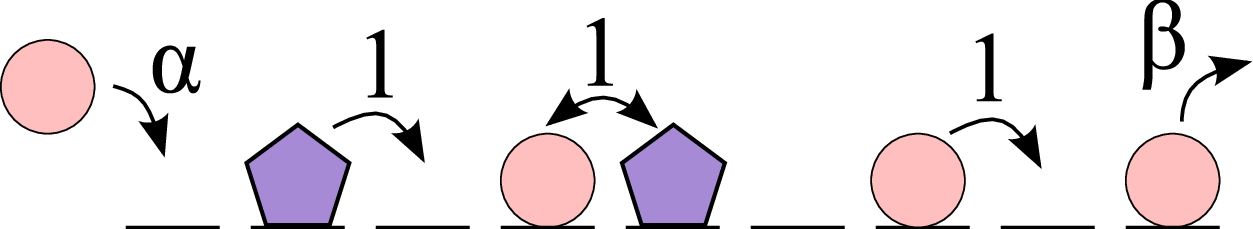}
\end{center}
\caption{\label{fig:2cp} Interactions between first class (circles) and second class (hexagons) particles. Second class particles hop can forward into empty sites and exchange places with first-class particles behind them. Note that in our simulations, at most one second-class particle was present at any instant, and that the behaviour at the boundaries was chosen so as not to affect the motion of first-class particles (see text).}
\end{figure}

The other observable we investigated was the position of a second class particle \cite{BCFG89,Ferrari91,DJLS93}.  Like a first-class particle, a second-class particle hops to the right into vacant sites.  However, it may also exchange places with a particle occupying the site to its left. Both of these processes take place at unit rate, and first-class hop to the right at the same rate, irrespective of whether the site in front of them is empty or occupied by a second-class particle---see Fig.~\ref{fig:2cp}.  So as not to affect the entry of first-class particles into the system, a second-class particle on the left-boundary site is forced to exit if a first-class particle attempts to enter.  It is reinserted as soon as the left-boundary site becomes vacant again.  Second-class particles are prevented from exiting at the right boundary.  In these simulations one can measure the position of the second class particle, $x(t)$, starting from an initial condition where each site is occupied by a first-class particle with density $\rho$ (as described above) except for the left boundary site, which is always initially occupied by the second-class particle.  Here, the analogue of $R(t)$ is
\begin{equation}
\xi(t) = \frac{\langle x(t) \rangle - \langle x \rangle}{x(0) - \langle x \rangle} \;.
\label{scp}
\end{equation}

\begin{figure}
\begin{center}
\includegraphics[width=0.66\linewidth]{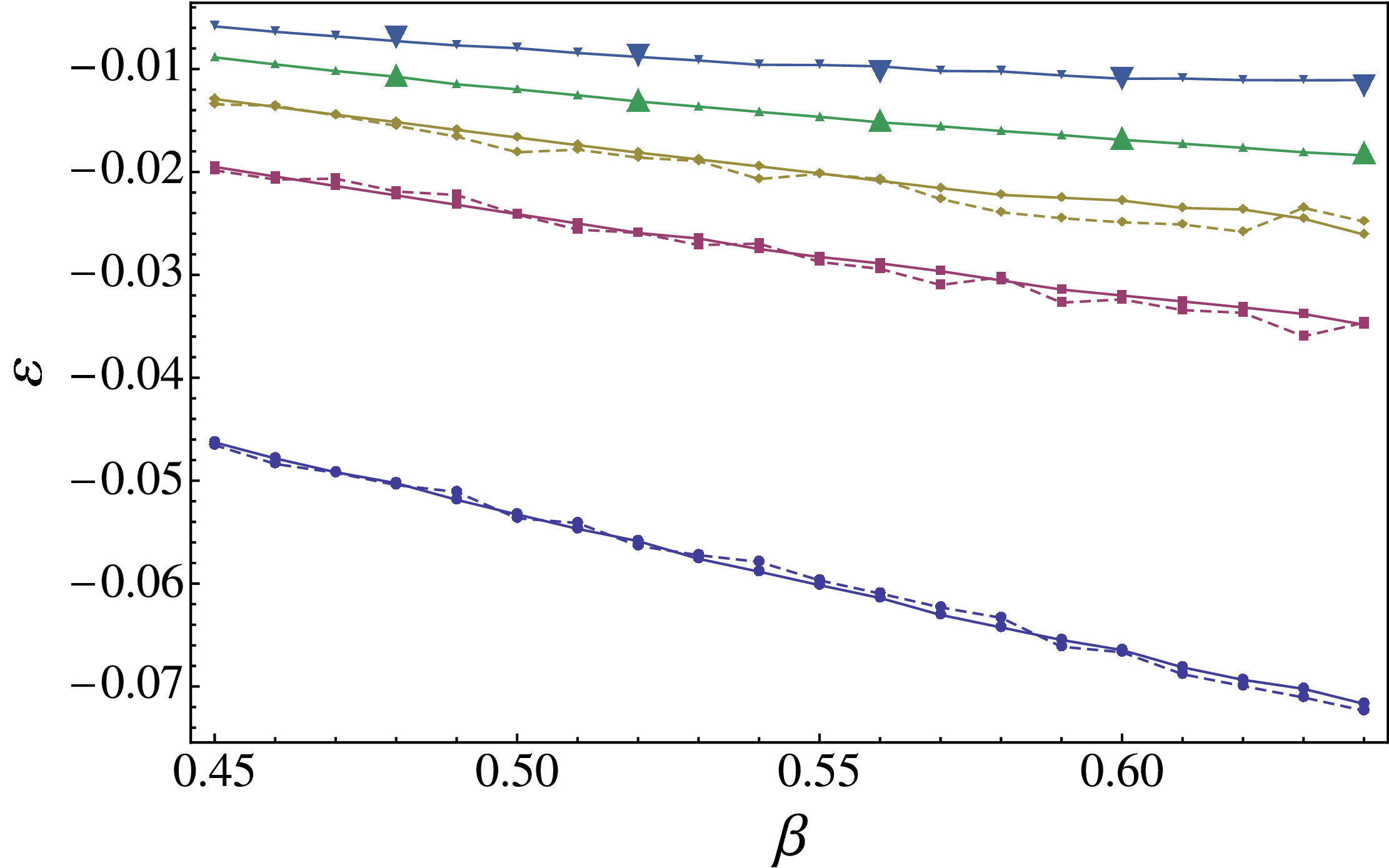}
\end{center}
\caption{\label{fig:observables} The gap as a function of $\beta$ as obtained by consideration of the relaxation of three different observables. Solid lines show data obtained from the relaxation of the density for system sizes $L=10,20,30,50,150$ (bottom to top).  Alongside the lower three curves are plotted data from the stationary density auto-correlation function (\ref{chi}) (dashed lines), and alongside the upper two data from the position of a single second-class particle (\ref{scp}) (large symbols). All three observables yield consistent results.}
\end{figure}

Gaps obtained from these two functions $\chi(t)$ and $\xi(t)$ are compared with those obtained from $R(t)$ in Fig.~\ref{fig:observables} at various system sizes.  We see that there are no systematic differences between these functions at the system sizes studied, and therefore that they are unlikely to provide improved estimates of the gap in the thermodynamic limit $L\to\infty$.  We therefore conclude that whilst Monte Carlo simulation data is possibly more consistent with the de~Gier--Essler prediction for the gap than the domain wall theory, the effect of the changing gap on the relaxation of a typical observable is very small and hard to disentangle from the noise.

\section{Evidence for the dynamical transition from density matrix renormalisation}
\label{sec:dmrg}

Given the difficulties in measuring the gap from Monte Carlo simulations, we now turn to a fundamentally different numerical approach, namely the density matrix renormalisation group (DMRG) \cite{DMRGBook} that was briefly discussed in the introduction.  We begin by recapitulating the essential features of the DMRG approach before demonstrating that it does indeed show a dynamical transition along the de~Gier--Essler line.

\subsection{DMRG algorithm for the TASEP gap}
\label{subsec:dmrg_algorithm}

The density matrix renormalisation group (DMRG) procedure builds an
approximation to the transition matrix $M$ that appears in the master
equation (\ref{eq:me}). The basic idea is to repeatedly add sites to the
system and renormalise the enlarged transition matrix so that its
dimensionality remains constant as the system size grows.  In this
way, the approximated transition matrix remains sufficiently small that its
spectrum (and in particular the gap) can be determined using standard
numerical diagonalisation methods. The success of this procedure relies on
finding a reduced basis set at each renormalisation step that allows the
lowest-lying eigenstates to remain well approximated as sites are added.

Since the TASEP has open boundaries, the extension of the lattice is achieved in this instance by adding sites to the \emph{middle} of the system rather than at one of the ends \cite{Nagy02}.  The system is thus divided into two `blocks': the left and right halves of the lattice. It is extended in each iteration by adding a site at the right-hand end of the left block, and the left-hand end of the right block.  In order to understand the results that come out of the procedure, it is worth explaining in a little more detail how the renormalisation is actually achieved---full details are provided in \ref{app:dmrg}.

\begin{figure}
\centering
\includegraphics[width=0.4\linewidth]{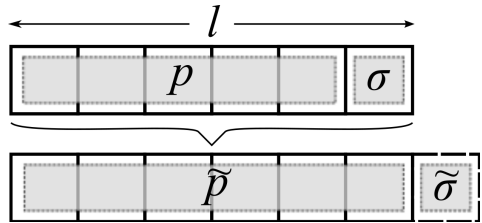}
\caption{\label{fig:dmrgstep} Illustration of the DMRG procedure applied to the left block. Before the transformation, the state of the block is specified by $1\le p\le m$ and $\sigma=0,1$. After adding a new site to the end of the block, the original $\ell$ sites are renormalised so that they are specified by the two numbers $1\le \tilde{p} \le m$ and $\tilde{\sigma}=0,1$.  The right block is a mirror image of the left, and treated the same way.}
\end{figure}

We outline the procedure in terms of the transformation applied to the left block.  Suppose that at the start of the iteration, the block comprises $\ell$ lattice sites.  The set of states the block can be in is specified by two `quantum' numbers, $p = 1,2 \ldots, m$ and $\sigma=0,1$.  The quantity $p$ indexes the configuration of the first $\ell-1$ sites, and $\sigma$ the occupancy of the rightmost site.  The first step of the renormalisation is to construct some basis in the $2m$-dimensional space spanned by $p$ and $\sigma$ and, crucially, retain only $m$ of these basis vectors.  This defines a renormalised  index $\tilde{p}=1,2,\ldots,m$ for the \emph{entire} $\ell$-site block.  A new site is then added to the right hand side of the lattice.  The configuration of this additional site is specified by the renormalised coordinate $\tilde{\sigma}$.  See  Fig.~\ref{fig:dmrgstep}. The reason for keeping the rightmost site in the `physical' (site occupancy) basis is that one needs to project the transition matrix for hops from site $\ell$ to $\ell+1$ onto the space spanned by $\tilde{p}$ and $\tilde{\sigma}$. These transition rates are initially only known in this physical basis. Thus through this procedure, we obtain a (truncated) representation of the transition matrix for an $\ell+1$-site system in a space of the same dimensionality as that used to describe the $\ell$-site system.  

The same transformation is applied to the right block, the only difference being that it is the leftmost site that is expressed in the physical basis.  Interactions between the two blocks enter through the construction of the renormalised indices $\tilde{p}$ as we now describe.  First, a transition matrix $M'$ for the entire system of $2\ell$ sites is constructed by combining the matrices for the two halves, and by adding an interaction term that allows a particle in the left block to hop to the right block.  Again, having the internal two sites specified in the physical basis helps here.  This transition matrix is then diagonalised.  However, it is not these eigenstates that are used to perform the renormalisation.  Rather, a \emph{density matrix}, which is a symmetric combination of the two longest-lived eigenstates of $M'$, is constructed, and eigenvectors of this density matrix are instead used for the renormalisation.  The idea is that this form of the density matrix allows the stationary and longest-lived transient state to be accurately represented in large systems.  The specific form of the density matrix, and the prescription for obtaining the truncated basis set, is given in \ref{app:dmrg}. For further details about the principles behind DMRG and other applications we refer to \cite{DMRGBook,White98,Noack05}, and especially \cite{Schollwock05}.

\subsection{DMRG results}
\label{subsec:dmrg_results}

We used the DMRG algorithm outlined above to estimate the gap for a given combination of $\alpha$ and $\beta$ by starting with an exact diagonalisation of the $L=8$ system, and keeping $m=16$ eigenvectors of the density matrix in each renormalisation step.  In principle, one ought to be able to access arbitrarily large system sizes with this method.  In practice---and as was also noted in \cite{Nagy02}---the algorithm eventually goes unstable, which is typically manifested through the gap acquiring an imaginary part.  We simply ignore data for system sizes where the instability is judged to have kicked in.

\begin{figure}
\centering
\includegraphics[width=0.66\linewidth]{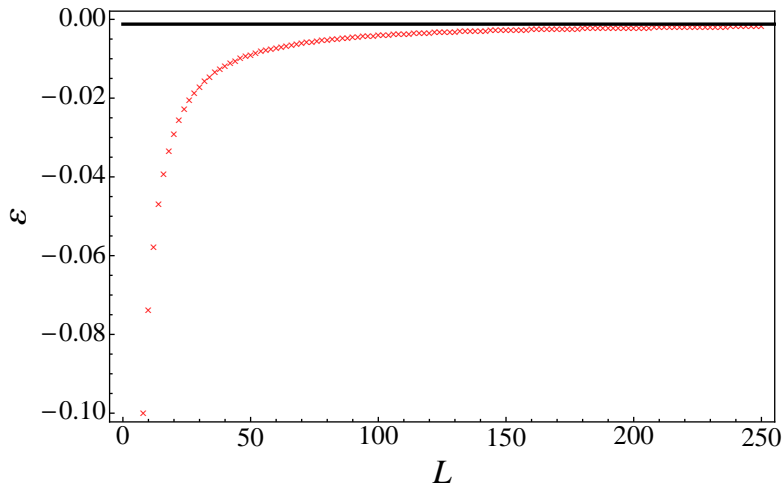}
\caption{Stable DMRG data for the gap (red crosses) and the exact thermodynamic gap (black line) at $\alpha = 0.4$ and $\beta = 0.75$.}
\label{fig:dmrgRawData}
\end{figure}

\begin{figure}
  \centering
  \subfigure[]{\label{fig:dmrgRsq}%
    \includegraphics[width=0.45\linewidth]{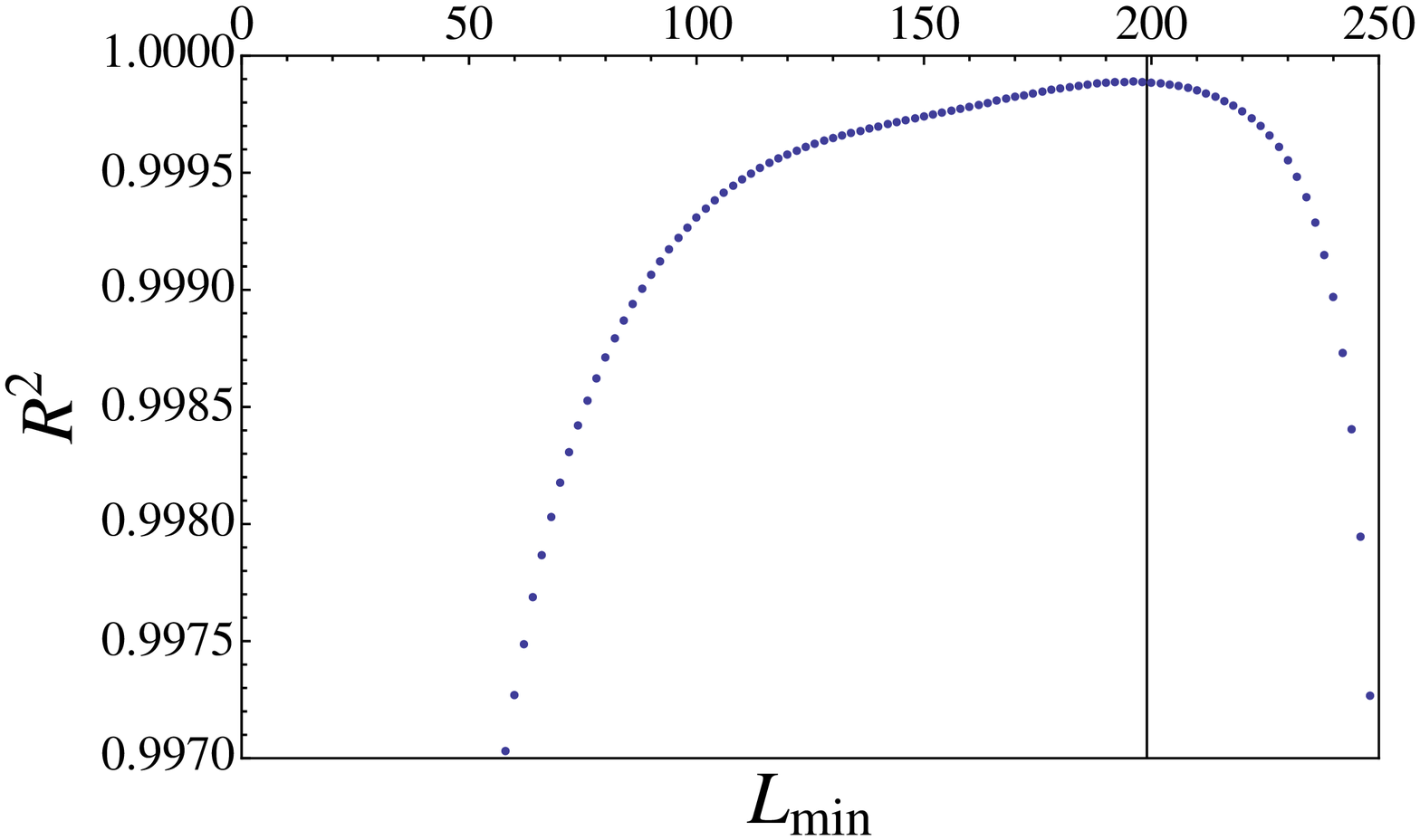} }
  \subfigure[]{\label{fig:dmrgFitGaps}%
    \includegraphics[width=0.45\linewidth]{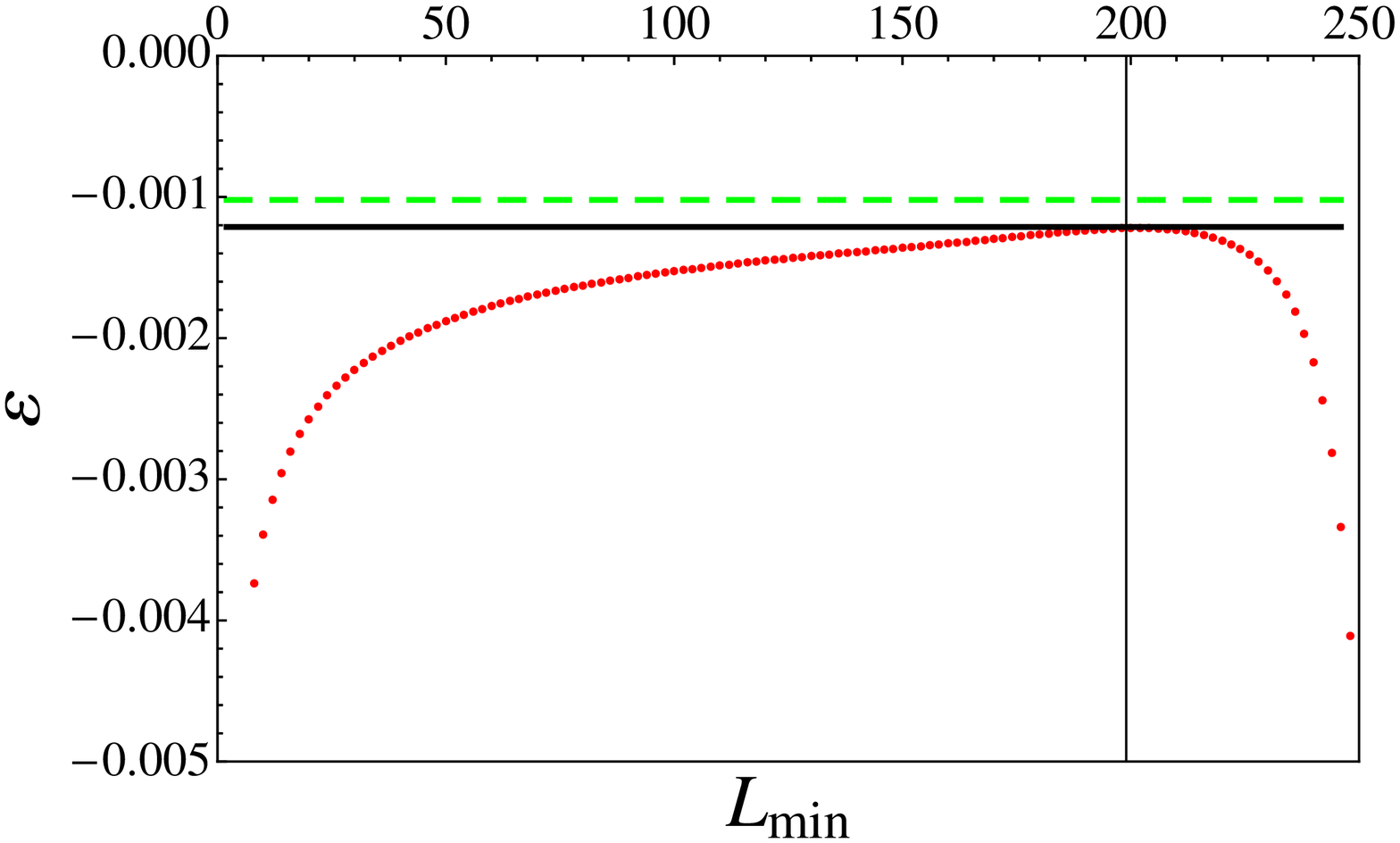} } 
  \caption{ \subref{fig:dmrgRsq} Coefficient of determination $R^2$ and \subref{fig:dmrgFitGaps} extrapolated gap, as a function of $L_{\rm min}$ for DMRG data corresponding to $\alpha = 0.4$, $\beta = 0.75$. The optimal choice of $L_{\rm min}$ lies just below $200$, for which the extrapolated gap very closely matches the de~Gier-Essler result and clearly differs from the DWT gap. }
\end{figure}

Although we can access larger system sizes with the DMRG approach than was possible with Monte Carlo (e.g., $L$ up to about 250, as shown in Fig.~\ref{fig:dmrgRawData}), it is still necessary to extrapolate to the thermodynamic limit, $L\to\infty$.  We follow a similar procedure to that described in Section~\ref{sec:decay}.  That is, we specify a finite-size fitting function of the form $f(L) = \varepsilon + a_2 L^{-2}$, and adjust the smallest value of $L$ used for the fit.  Again, we use ${\cal R}^2$ as a goodness-of-fit measure, i.e., Eq.~(\ref{R2}) with $t$ replaced by $L$.  We show in Fig.~\ref{fig:dmrgRsq} how the goodness of fit varies with the minimum value of $L$ used in the fit; choosing the optimal (largest) value yields an estimate of the gap that we observe from Fig.~\ref{fig:dmrgFitGaps} that appears more consistent with the de~Gier--Essler prediction than domain wall theory.

\begin{figure}
\centering
\includegraphics[width=0.66\linewidth]{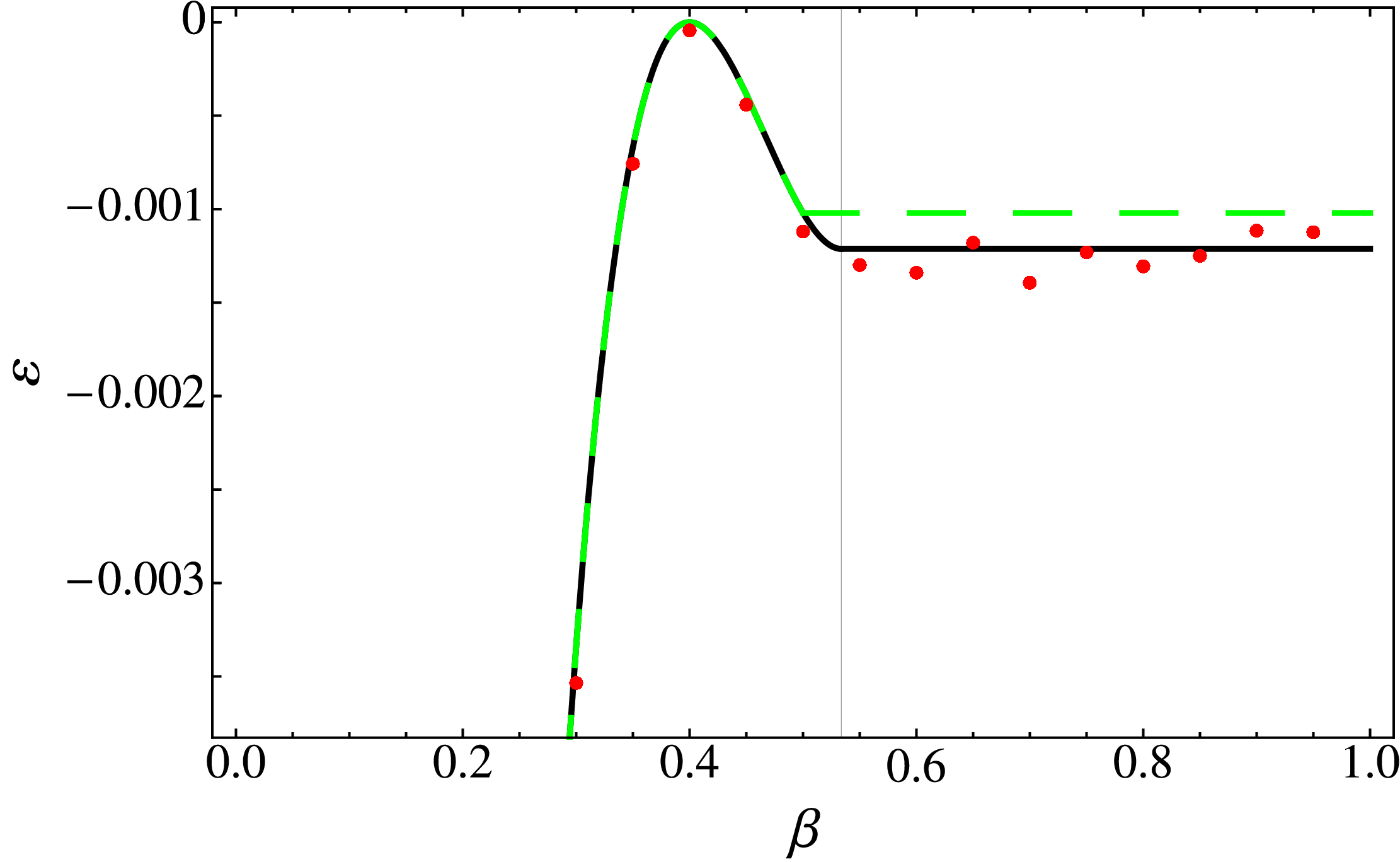}
\caption{DMRG estimates of the gap (red crosses), exact thermodynamic gap (black curve) and DWT gap (green curve), all as a function of $\beta$ at $\alpha = 0.4$ (for which $\beta_c \approx 0.53$).}
\label{fig:dmrggapacr}
\end{figure}

We show the DMRG estimate of the gap obtained in this way as a function of $\beta$ in Fig.~\ref{fig:dmrggapacr}.  The agreement with the analytical prediction common to dGE and DWT below $\beta < \frac{1}{2}$ is very good.  For larger $\beta$ the data are scattered around an apparently constant value, indicating the presence of inaccuracies in the DMRG algorithm and/or the extrapolation of $L\to\infty$.  If we regard these data as independent measurements of the same value, we can obtain an estimate of the uncertainty in the constant value of the gap above some critical point by simply calculating the standard deviation.  We find the DMRG gap to approach the constant value $\varepsilon = -0.00125(9)$ when $\alpha=0.4$.  For comparison, de Gier and Essler predict a constant value for the gap of $-0.00121\ldots$ above the dynamical transition, while domain wall theory predicts $-0.00102\ldots$.  The DMRG measurement is then clearly consistent with the de~Gier--Essler prediction, whilst the difference from the domain wall theory value is significant.  Taking this result together with the Monte Carlo data we conclude that a dynamical transition does indeed occur at the dGE line, and not where DWT predicts.

\section{Discussion of the distinction between static and dynamic subphases}
\label{sec:discuss}

In this paper we have presented numerical evidence, the most convincing of which came from DMRG
calculations, that a dynamical transition occurs in the TASEP at the dGE line rather than at the subphase boundary.  However, this finding in turn raises a number of open questions. For example, does the dGE line---and in particular its
departure from the static subphase boundary---have any physical significance?  How is the static subphase boundary manifested in the eigenvalue spectrum?  We discuss the distinction between the two phase diagrams first in general theoretical terms, and then with reference to different treatments of the TASEP.

\subsection{General theory of nonequilibrium phase transitions}

One way to gain a general understanding of nonequilibrium phase transitions is through the analogue of partition function for a system of size $L$, $Z_L$ \cite{BE02,Altenberg02}.  It has been shown that for any system governed by Markovian dynamics, it may be written as
\begin{equation}
Z_L= \prod_{j\neq0} (-\lambda_j)
\label{Z}
\end{equation}
where $(-\lambda_j)$ are the eigenvalues of the Markov transition matrix which defines the dynamics \cite{Blythe01phd,Altenberg02}. For a finite, irreducible configuration space there is a unique stationary state corresponding to
the largest eigenvalue $\lambda_0=0$. By the Perron-Frobenious theorem \cite{Senata}, $\lambda_0$
is non-degenerate, therefore the gap is given by the second largest eigenvalue $\lambda_1$
and the largest relaxation time is $\tau_1 = -1/\lambda_1$.

From our knowledge of equilibrium phase transitions, we expect a
static phase transition to occur when $\lim_{L\to\infty}\frac{1}{L}\ln
Z_L$ exhibits nonanalyticities.  We see from (\ref{Z}) that to obtain a
static phase transition we require a major restructuring of the
eigenvalue spectrum.  To be general and explicit let us consider the
nonequilibrium analogue of the free energy density,
\begin{equation}
h_L = \frac{\ln Z_L}{L} = \frac{1}{L} \sum_{\lambda_i \ne 0} \ln (- \lambda_i )  \;.
\label{h}
\end{equation}
Consider an arbitrary eigenvalue $\lambda_i$ in
(\ref{h}), possessing a nonanalyticity at a
critical value of some parameter. Unless a significant
number of other eigenvalues in the spectrum 
converge onto the same nonanalyticity in the thermodynamic limit, the effect of the
$\lambda_i$ nonanalyticity is lost as $L \to \infty$ and
(\ref{h}) and hence $h_L$ remain analytic.

On the other hand a dynamical phase transition, as defined above, is
only concerned with the eigenvalue $\lambda_1$. Therefore a dynamical
phase transition does not necessarily coincide with a static phase
transition.  One very simple example would be if the two leading
subdominant eigenvalues $\lambda_1$ and $\lambda_2$ crossed at some
values of a control parameter. Then the gap would behave in a nonanalytic
way but $Z$ would be analytic.

Let us consider more explicitly the TASEP. From the exact solution of the TASEP \cite{DEHP93,BE07}
the expression for $Z_L$, as defined by (\ref{Z}), is
\begin{equation}
Z_L =  \frac{(\alpha \beta)^{L+1}}{\beta-\alpha}\left[ \Phi_L(\alpha) - \Phi_L(\beta)\right]
\label{Z2}
\end{equation}
where
\begin{equation}
\Phi_L(x)= \sum_{p=1}^L \frac{p (2L-p-1)!}{L! (L-p)!}  x^{-(p+1)}
\end{equation}
and for $L$ large
\begin{equation}
\Phi_L(x)
\simeq \left\{ \begin{array}{ll}
\displaystyle \frac{x (1-2x)}{[x(1-x)]^{L+1}} & x < \frac{1}{2} \\[2ex]
\displaystyle \frac{4^L}{\sqrt{\pi L}} & x = \frac{1}{2} \\
\displaystyle \frac{x\, 4^L}{\sqrt{\pi} (2x-1)^2 L^{3/2}} & x >
\frac{1}{2}
\end{array}\right. \;.
\label{Phi}
\end{equation}

Now consider, for example, the case $\alpha <1/2$. As we increase
$\beta$ a first-order static phase transition occurs on the line
$\beta =\alpha $ where the dominant contribution to (\ref{Z2}) changes
from $\Phi_L(\beta)$ to $\Phi_L(\alpha)$.
Thus at the phase transition $h_L$ defined in (\ref{h}) is non-analytic.
The subphase boundary is at
$\beta = 1/2$ where the asymptotic behaviour of the subdominant
contribution, $\Phi_L(\beta)$ changes.
As this is a subdominant contribution to $\ln Z_L$,
$h_L$ remains  analytic at the sub-phase boundary.
At the dGE line  there is no noticeable
change in the asymptotic behaviour of $Z_L$
and $h_L$ is analytic.

\subsection{Static vs dynamic transitions within various treatments of the TASEP}

We illustrate further the distinctions between the static and dynamics phase diagrams
by collecting exact results for the TASEP together with the predictions of
various level of approximations such as mean field theory and domain
wall theory.

\paragraph{Burgers Equation (Mean Field theory)}
We begin by considering a mean field description of the system
that is given by the Burgers equation,
\begin{equation}
\label{eq:burg}
  \partial_t\rho(x,t) = -(1-2\rho)\partial_x\rho + \frac{1}{2}\partial^2_x\rho 
\;,
\end{equation}
where $\rho(x,t)$ is a density field.
Although the Burgers equation has been
studied extensively in the literature, we are not aware of any
treatment of the case with prescribed boundary reservoir densities.
In the appendix we give the solution of (\ref{eq:burg}) subject to the
boundary conditions $\rho(0) = \alpha$ and $\rho(L) = 1-\beta$, and
arbitrary initial condition.  

\begin{figure}
  \centering
  \includegraphics[width=0.6\linewidth]{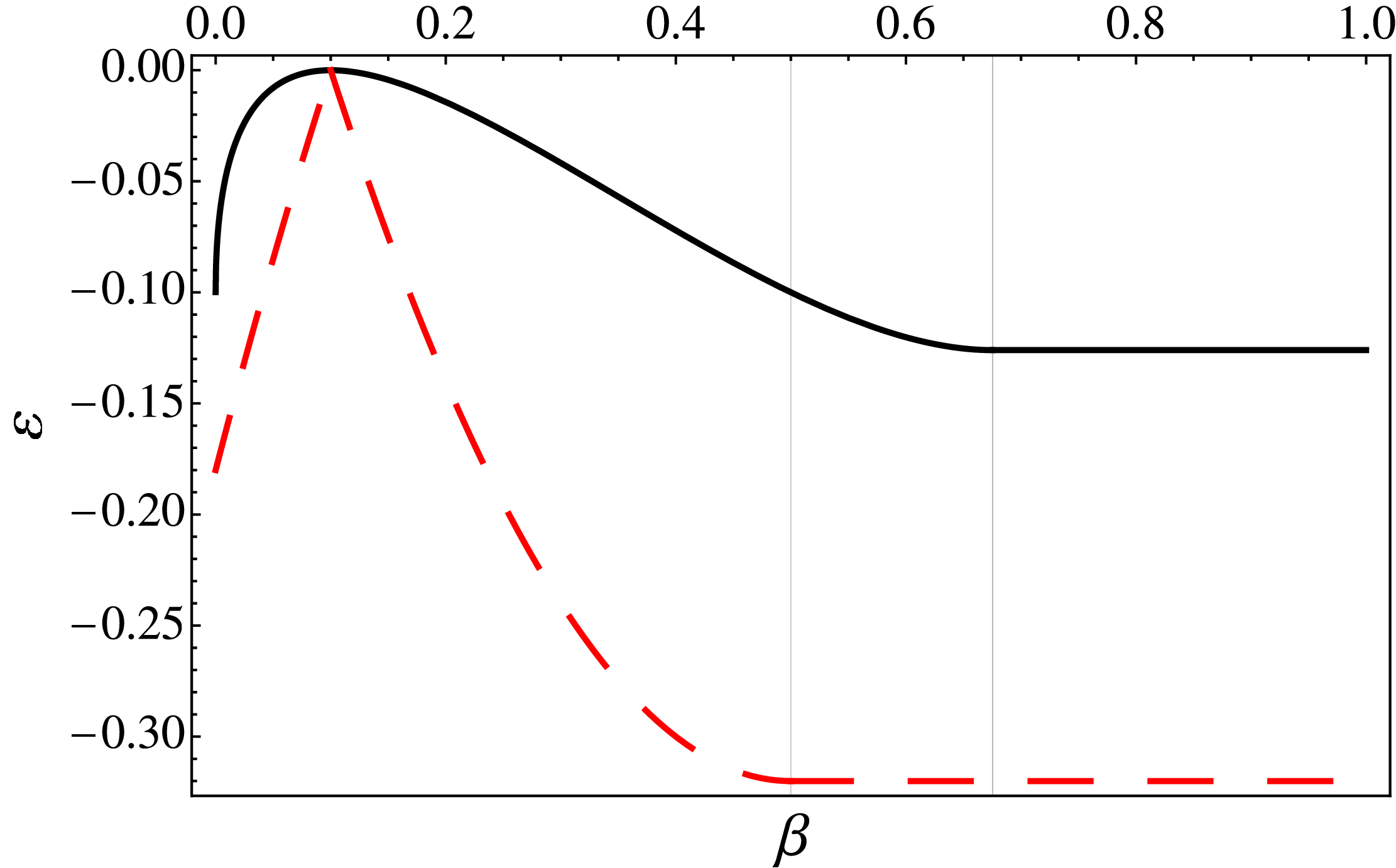} 
  \caption{Comparison of the gap for $L \to \infty$ from the spectrum of the viscous Burgers equation (\ref{eq:burg}) to the exact thermodynamic gap, for $\alpha = 0.1$ and $\beta_c \approx 0.675$.}
  \label{fig:GapComparisonBurgersExact}
  \label{fig:bgap}
\end{figure}

The results for the Burgers gap $\varepsilon_B$ may be summarised as follows
(restricting ourselves to the case $\alpha <1/2$):
\begin{eqnarray}
\mbox{For}\quad \beta <1/2\quad
\varepsilon_B &=& -2 | \alpha(1-\alpha) - \beta(1-\beta) | \\
\mbox{For}\quad \beta >1/2\quad
\varepsilon_B &=& -  \frac{(1-2\alpha)^2}{2}
\end{eqnarray}
The behaviour of the gap is plotted in Fig.~\ref{fig:bgap}.  Note that
a dynamical transition occurs at $\beta = 1/2$.  Interestingly, it turns out
that there is no change in the density profile at this value so, in
fact, the mean field theory predicts a dynamical transition instead of
a subphase boundary at $\beta = 1/2$.
Thus even at the level of mean-field theory, the TASEP exhibits the phenomenon of a dynamic transition that is not accompanied by a static transition.

\paragraph{Domain wall theory}  The predictions of domain wall theory were reviewed in Section~\ref{subsec:DWT}.  As we pointed out there, DWT gives a subphase boundary at $\beta=1/2$ along with a dynamical transition: the distinction between the two phase diagrams is lost.  However, in the subphases LDI and HDI, the DWT gives the correct dynamical gap to order $1/L^2$.

\paragraph{Exact results}  As was discussed in Section~\ref{sec:moddef}, the
exact results for the stationary state \cite{DEHP93,Schutz93} 
and the spectrum \cite{deGier05,deGier06}
reveal behaviour that is different from both mean-field and domain wall theory. In common
with the latter, there is a static subphase boundary at $\beta = \frac{1}{2}$. Like the mean field
theory, the location of the dynamic transition occurs at a distinct location to the static phase boundary.  However, this location $\alpha_c(\beta)$ is nontrivial and is not predicted by either of the simpler theories.  Moreover, the behaviour of the gap at the static transition point $\beta=\alpha$ differs between the mean-field and exact theories: in the former, the gap has a cusp, whereas in the latter it varies analytically.

\section{Conclusion}
\label{sec:conclude}

In this work we have provided numerical evidence to 
confirm the existence of the dynamical transition line
predicted by de Gier and Essler. 
Ultimately the evidence came
from the approximate, but quantitatively reliable, DMRG technique.

To conclude, we now return to the main open question resulting from this work,
namely that of the physical significance of the dynamical transition that
takes place along the de~Gier--Essler line.  We  remark again that, as was
seen in Section~\ref{sec:sims}, direct measurement of the gap in Monte Carlo
simulations was very difficult. The nonanalyticity in the leading eigenvalue
was barely detectable in the stochastic simulation data.  We were therefore
unable to identify, for example, whether the system relaxed to stationarity
in a fundamentally different way either side of the transition line.

Given the success of the domain wall theory in predicting the gap in part of the phase diagram,
it is tempting to believe that with some refinement it may
be able to explain the dynamical transition and furthermore reveal the physics associated
with it. For example, it has been noted in\cite{deGier06}, that DWT may be modified to
give the correct gap throughout the LD1$'$ region by simply imposing that the right
domain density be $\rho^+ = 1-\beta_c$ in this region.
However DWT then no longer predicts a static subphase boundary at $\beta=1/2$
and also the transition line $\alpha_c(\beta)$ is not predicted.

It may be that some knowledge of the form of the eigenvectors associated to the low lying states will allow an understanding of what occurs at the dynamical transition.  
It would be of interest to extend the DMRG studies to investigate the
form of the eigenfunctions. Also recent analytic work by Simon \cite{Simon09},
in which eigenvectors are constructed for the partially asymmetric exclusion process
through the co-ordinate Bethe ansatz, may help to shed light on the matter.

\ack

R.A.B.\ thanks the RCUK for the support of an Academic Fellowship. AP would like to thank Fabian Essler for hospitality and discussion,
and acknowledges financial support from the EPSRC.\\

\appendix
\section{DMRG algorithm for TASEP}
\label{app:dmrg}

In this appendix, we provide a step-by-step description of the DMRG algorithm used
to obtain the results presented in Section~\ref{sec:dmrg}.

In what follows we make use of three elementary transition matrices for the TASEP. With the ordering $(\0, \1)$ of single-site configurations we define
\begin{equation}
  h_l = 
\left(  \begin{array}{cc}
    -\alpha & 0\\
    \alpha & 0
  \end{array}\right) \quad\mbox{and}\quad
  h_r = 
\left(  \begin{array}{cc}
    0 & \beta\\
    0 & -\beta
  \end{array}\right)
\end{equation}
for the entry and exit of particles at the left and right boundary sites.  In the bulk, and with the ordering $(\0\0, \0\1, \1\0, \1\1)$ of two-site configurations, we define
\begin{equation}
  h_b = 
\left(  \begin{array}{cccc}
    0 & 0 & 0 & 0\\
    0 & 0 & 1 & 0\\
    0 & 0 & -1 & 0\\
    0 & 0 & 0 & 0
  \end{array} \right) \;.
\end{equation}

The DMRG algorithm now proceeds as follows.

\newcounter{dmrgstep}
\begin{list}{\arabic{dmrgstep}.}{\usecounter{dmrgstep}\setcounter{dmrgstep}{-1}}
\item \textit{Initialise block transition matrices}\quad
We partition the system into two blocks, each of size $\ell$.
The transition matrix for the left block can be written as
\begin{equation}
  M_L^{(\ell)} = h_l \otimes \I^{\otimes \ell-1} +  \sum_{k = 1}^{\ell-1} \I^{\otimes k-1} \otimes h_b \otimes \I^{\otimes \ell-k-1} \;,
\end{equation}
where $\I$ is the $2\times2$ identity matrix.  Likewise, for the right block one has
\begin{equation}
  M_R^{(\ell)} = \sum_{k = 1}^{\ell-1} \I^{\otimes k-1} \otimes h_b \otimes \I^{\otimes l-k-1} + \I^{\otimes\ell-1} \otimes h_r \;.
\end{equation}
\item \textit{Diagonalise the transition matrix for the entire system}\quad 
The transition matrix for the entire system of size $2\ell$, $M^{(2\ell)}$, is given by
\begin{equation}
\label{M2l}
M^{(2\ell)} = M_L^{(\ell)} \otimes \I^{\otimes \ell} +  \I^{\otimes \ell - 1} \otimes h_b \otimes \I^{\ell - 1} + \I^{\ell} \otimes M_R^{(\ell)} \;.
\end{equation}

Now we diagonalise $M^{(2\ell)}$, first using the Arnoldi algorithm \cite{Golub1996} (as implemented in the ARPACK package \cite{ARPACK} and accessed using Mathematica) to find the left and right ground state eigenvectors $\langle \phi_0|$, $|\psi_0\rangle$. We also want
to compute the gap and associated eigenvectors $|\psi_1\rangle$ and $\langle \phi_1|$.  Following \cite{Nagy02}, we find this can be done more accurately by `shifting' the transition matrix through $M' = M^{(2\ell)} + \Delta |\psi_0\rangle\langle\phi_0|$ sufficiently far that the gap is now the largest eigenvector of $M'$.
\item \textit{Form reduced density matrices}\quad
The reduced density matrices are defined as
\begin{eqnarray}
  \rho_L &= \frac{1}{4} \Tr_R(|\psi_0\rangle\langle\psi_0| + |\phi_0\rangle\langle\phi_0| + |\psi_1\rangle\langle\psi_1| + |\phi_1\rangle\langle\phi_1| ) \\
  \rho_R &= \frac{1}{4} \Tr_L(|\psi_0\rangle\langle\psi_0| + |\phi_0\rangle\langle\phi_0| + |\psi_1\rangle\langle\psi_1| + |\phi_1\rangle\langle\phi_1| )  \;.
\end{eqnarray}
Here $\Tr_{L}$ and $\Tr_{R}$ indicate a trace over the degrees of freedom in the left and right blocks respectively.

To be clear, let $\{ |i\rangle_{L} \}$ and $\{ |j\rangle_{R} \}$ be basis sets for the left and right blocks respectively.  A state vector $|\psi\rangle$ across both blocks can then be expanded as
\begin{equation}
  |\psi\rangle = \sum_{ij} c_{ij} |i\rangle_L |j\rangle_R \;.
\end{equation}
The trace operation $\Tr_L(|\psi\rangle \langle \psi|)$ is then defined as
\begin{equation}
\Tr_L(|\psi\rangle \langle \psi|) = \sum_{j j'} \left(\sum_{i} c_{ij} c_{ij'}\right) | j\rangle_R \langle j' |_{R} \;.
\end{equation}
Note that this is an operator acting only on the right block.  The corresponding operation $\Tr_R$ is defined analogously.
\item \textit{Diagonalise the density matrices and form a truncated basis set}\quad
We first find all eigenvalues and eigenvectors of the symmetric matrices $\rho_L$ and $\rho_R$ using a dense matrix diagonalisation routine from the LAPACK package \cite{LAPACK}, again accessed using Mathematica. We then form the matrices $O_L$ and $O_R$ that have as their columns the $m$ eigenvectors with largest eigenvalues of $\rho_L$ and $\rho_R$ respectively. Then we construct projectors
\begin{equation}
\label{red}
  P_L = O_L \otimes \I \quad\mbox{and}\quad
  P_R = \I \otimes O_R
\end{equation}
that will be used in the renormalisation step below.
\item\textit{Enlarge system}\quad  The left block is enlarged by adding a site at its
right end.  The transition matrix for this enlarged block is
\begin{equation}
  \widetilde{M}_L^{(\ell+1)} = M_L^{(\ell)} \otimes \I^{\otimes 2}  +  \I^{\otimes \ell-1} \otimes h_b \;.
\end{equation}
Likewise, the right block is enlarged by adding a site at its left:
\begin{equation}
  \widetilde{M}_R^{(\ell+1)} = h_b \otimes \I^{\ell-1}  +  \I^{\otimes 2} \otimes M_R^{(\ell)}.
\end{equation}
\item\textit{Renormalise the transition matrices for the enlarged system}\quad
The renormalisation is performed by projection the enlarged transition matrices onto the
subset of density matrix eigenstates that was retained in step 3.  Formally, this is
achieved by computing
\begin{equation}
  M_L^{(\ell+1)} = P_L^T \widetilde{M}_L^{(\ell+1)} P_L \quad\mbox{and}\quad
  H_R^{(\ell+1)} = P_R^T \widetilde{M}_R^{(\ell+1)} P_R  \;.
\end{equation}
Note that after this transformation, the first $\ell$ sites of the system are no longer represented in the `physical' basis (i.e., particle-hole configurations).  However, the form of the projectors (\ref{red}) ensures that the last site of the block \emph{is} represented in the physical basis, which is essential for the expression (\ref{M2l}) to be valid at each stage of the renormalisation.
\item\textit{Return to step 1}, putting $\ell \to \ell+1$.
\end{list}

\section{Solution of the Burgers equation with fixed boundary densities}

Here we solve the Burgers equation (i.e., the continuum limit of the TASEP within a mean-field theory) 
\begin{equation}
\label{eq:burg2}
  \partial_t\rho(x,t) = -(1-2\rho)\partial_x\rho + \frac{1}{2}\partial^2_x\rho \;.
\end{equation}
subject to the boundary conditions $\rho(0) = \alpha$ and $\rho(L) = 1-\beta$.
We use  the Cole-Hopf transformation
\cite{Woyc98}, which involves the change of variable
\begin{equation}
\rho(x,t) = \frac{1}{2}\left[ 1 + \frac{\partial}{\partial x} \ln u(x,t) \right] \;.
\end{equation}
Substituting into (\ref{eq:burg2}), and integrating with respect to $x$
eventually yields
\begin{equation}
\label{eq:ch}
\frac{\partial}{\partial t} u(x,t) = \frac{1}{2} \frac{\partial^2}{\partial x^2} u(x,t) \;.
\end{equation}
The boundary conditions on $\rho(x,t)$ transform to the conditions
\begin{eqnarray}
\label{eq:ubc1}
u'(0,t) &=& (2\alpha-1) u(0,t) \\
\label{eq:ubc2}
u'(L,t) &=& (1-2\beta) u(L,t)\;.
\end{eqnarray}
We proceed by finding the eigenfunctions $\phi_n(x)$  that satisfy
\begin{equation}
\label{phin}
\frac{1}{2} \frac{{\rm d}^2}{{\rm d} x^2} \phi_n(x) =  \lambda_n \phi_n(x)
\end{equation}
along with the above boundary conditions at $x=0$ and $x=L$.  Thus
given an initial condition $\rho(x,0)$ the function $u(x,t)$ will be
given by
\begin{equation}
u(x,t) = \sum_{n\ge 0} c_n {\rm e}^{\lambda_n t} \phi_n(x)
\end{equation}
where the coefficients $c_n$ will depend on the initial condition.  Inverting the Cole-Hopf transformation gives
\begin{equation}
\label{eq:ich}
\rho(x,t) = \frac{1}{2} \left[ 1 + \frac{\phi_0'(x)}{\phi_0(x)} + \frac{ \sum_{n>0} c_n {\rm e}^{(\lambda_n-\lambda_0)t} \phi_0(x) \left( \frac{\phi_n(x)}{\phi_0(x)} \right)' }{ \sum_{n \ge 0} c_n {\rm e}^{(\lambda_n-\lambda_0)t} \phi_n(x)} \right] \;.
\end{equation}
The final term, involving the ratio of sums, vanishes as $t\to\infty$, as all $\lambda_n$ with $n>0$ exceed $\lambda_0$.  Thus the stationary profile is
\begin{equation}
\rho(x) = \frac{1}{2} \left[1 + \frac{\phi_0'(x)}{\phi_0(x)}\right] \;.
\end{equation}

We can define the Burgers gap, $\varepsilon_B$ via the asymptotic rate of the exponential decay of the density profile to its stationary form, viz,
\begin{equation}
\varepsilon_B = \lim_{t \to \infty} \frac{{\rm d}}{{\rm d} t} \ln [\rho(x,t) - \rho(x)]  = \lambda_1 - \lambda_0 \;.
\end{equation}

The eigenfunctions given by (\ref{phin}) themselves take the form
\begin{equation}
\phi(x) = A {\rm e}^{\gamma x} + B {\rm e}^{-\gamma x}
\end{equation}
whence $\lambda = \frac{1}{2} \gamma^2$. The allowed values of $\gamma$ are quantised due to the boundary conditions (\ref{eq:ubc1}) and (\ref{eq:ubc2}). We first look for solutions with $\lambda>0$, i.e., real $\gamma$. Imposing the boundary conditions leads to the pair of equations
\begin{eqnarray}
\gamma (A  - B) &=& (2\alpha-1) ( A + B ) \\
\gamma ( A {\rm e}^{\gamma L} - B {\rm e}^{-\gamma L} ) &=& (1-2\beta) ( A {\rm e}^{\gamma x} + B {\rm e}^{-\gamma x} ) \;.
\end{eqnarray}
These equations are consistent only if
\begin{equation}
\tanh (\gamma L) = \frac{2(1-\alpha-\beta)\gamma}{(1-2\alpha)(1-2\beta)+\gamma^2} \;.
\end{equation}
In the limit $L\to\infty$, the left-hand side approaches $\pm 1$. The resulting equation has a solution $\gamma = 1-2\alpha$ if $\alpha<\frac{1}{2}$ and a solution $\gamma = 1-2\beta$ if $\beta<\frac{1}{2}$.

Solutions with a negative $\lambda$ have purely imaginary $\gamma = ik$ ($k$ real), and by following through the same procedure, one finds that the allowed values of $k$ are the roots of
\begin{equation}
\tan (k L) = \frac{2(\alpha+ \beta -1)k}{k^2-(2\alpha-1)(2\beta-1)} \;.
\end{equation}
In the large-$L$ limit, we find that $k \sim 1/L$, and hence the associated eigenvalues vanish as $1/L^2$ in the thermodynamic limit.

We can now state the $L\to\infty$ behaviour of the Burgers gap $\varepsilon$ as $\alpha$ and $\beta$ are varied.  When both $\alpha$ and $\beta$ are smaller than $\frac{1}{2}$, there are two isolated positive eigenvalues 
$\lambda = \frac{(2\alpha-1)^2}{2}, \frac{(2\beta-1)^2}{2}$.  In this region, then, the gap behaves as
\begin{equation}
\varepsilon_B = - \frac{|(2\alpha-1)^2 - (2\beta-1)^2|}{2} = -2 | \alpha(1-\alpha) - \beta(1-\beta) | \;.
\end{equation}
Like the exact gap (\ref{eq:gap_MI}), this vanishes along the HD-LD coexistence line $\alpha=\beta<\frac{1}{2}$. However, unlike the exact gap, the mean-field gap has a  nonanalyticity along this line.

If $\alpha$ or $\beta$ is increased above $\frac{1}{2}$, one of the two solutions with positive $\lambda$ ceases to exist, and the second-largest eigenvalue $\lambda_1=0$ in the thermodynamic limit. Hence then, the size of the gap is simply equal to the value of the largest eigenvalue. That is, within the low-density phase $\alpha<\frac{1}{2}, \beta>\frac{1}{2}$,
\begin{equation}
\varepsilon_B = -\frac{(1-2\alpha)^2}{2}
\end{equation}
and within the high-density phase $\alpha>\frac{1}{2}, \beta<\frac{1}{2}$ 
\begin{equation}
\varepsilon_B = -\frac{(1-2\beta)^2}{2} \;.
\end{equation}
The Burgers gap thus exhibits nonanalyticities along the coexistence
line $\alpha=\beta<\frac{1}{2}$, and along the lines
$\alpha<\frac{1}{2}, \beta=\frac{1}{2}$ and $\alpha=\frac{1}{2},
\beta<\frac{1}{2}$.
Thus  the dynamic phase diagram for the Burgers
equation appears the same as the static phase diagram, Fig.~\ref{fig:spd},
except that the subphase boundaries have become dynamical transition lines.

\end{document}